\documentclass[12pt,a4paper]{scrartcl}
\usepackage{graphicx} % Required for inserting images
\usepackage[utf8]{inputenc}
\usepackage{hyperref}
\usepackage[numbers,sort&compress]{natbib}
\usepackage{braket}
\usepackage{amsmath}
\usepackage{amsfonts}
\addtokomafont{caption}{\small}
\addtokomafont{captionlabel}{\bfseries}
\setcapindent{0pt}
\usepackage{amssymb}
\usepackage{graphicx}
\usepackage{float}
\usepackage{amsmath}
\usepackage{pdfpages}
\usepackage{caption}
\usepackage{subcaption}
\usepackage[rightcaption]{sidecap}
\usepackage[title]{appendix}
\usepackage{bm}

\usepackage{xcolor}

\def\p{\mathbf{P}}

\DeclareMathOperator{\sign}{sign}

\title{Chern numbers in quantum  graphs}

\author{Or Swartzberg\thanks{The Racah Institute of Physics, The Hebrew University of Jerusalem, Jerusalem, Israel 919040}, Omri Gat\footnotemark[1], and Boris Gutkin\thanks{School of Mathematical Sciences, Holon Institute of Technology, Holon 5810201, Israel} }

\begin{document}

\maketitle
\begin{abstract}\noindent 
    Quantum graphs provide an analytically tractable setting for the study of Chern numbers and band degeneracies in periodic systems. We study the Chern numbers  of energy bands  in a two-dimensional square lattice quantum graph.  We approach the problem by mapping the lattice  to a single-vertex quantum graph with two loops of equal lengths pierced by  magnetic fluxes. By establishing the degeneracy condition for its energy levels, we show that  
the model possesses two topological phases: a trivial phase, where the Chern numbers of all energy bands are $0$, and the nontrivial one, where the Chern numbers of successive energy bands alternate between $\pm1$. By applying the degeneracy condition, we calculate Chern-number phase diagrams analytically as a function of the node scattering matrix parameters and compare the results with numerical calculations. 
\end{abstract}
\tableofcontents

\section{Introduction}
In 1946, S.\ S.\ Chern introduced the topological invariants of complex vector bundles on smooth manifolds now known as Chern classes \cite{Chern1946}. Chern classes are derived from the coefficients of the characteristic polynomial of the curvature form associated with a vector bundle. In particular, the coefficient of the linear term in this polynomial is called the first Chern class \cite{Nakahara2003}, and the integral of this class over a two-dimensional base manifold is an integer known as the Chern number.

For parameter-dependent families of Hamiltonians,  each nondegenerate energy band can be naturally associated with a line bundle. Specifically,   let $H(\p)$ be  a  Hamiltonian at a point  $\p$ in the parameter  space manifold   $\mathcal{M}$ with energy bands $E_n(\p)$.
If the system is prepared in an eigenstate $|\psi_n(\p)\rangle $ of $n$th energy band , and the parameters are changed slowly enough along a closed one-dimensional path $\mathcal{C}\subset\mathcal{M}$, the final state, $e^{i\Delta\Phi}| \psi_n(\p)\rangle $, acquires the geometric phase   $\Delta\Phi =\int_{\mathcal{C}} A_n$, where $A_n=i\langle\psi_n|d \psi_n\rangle $ is the adiabatic (Berry) connection on the corresponding  line bundle \cite{Berry1984}. The adiabatic (Berry) curvature is $\Omega_n=dA_n$, and the first Chern class of the line bundle is the cohomology class $-\Omega_n/(2\pi)$; it follows that when $\mathcal{M}$ is two-dimensional, the Chern number $C_n$ of the $n$th energy band is $-\frac{1}{2\pi}\int_{ \mathcal{M}}\Omega_n$.

The topological invariance of the Chern numbers implies that $C_n$ remains constant when the Hamiltonian family is smoothly deformed as long as the gaps between the $n$th energy band and its neighbors remain open. In systems without time reversal symmetry, level crossings are generically codimension-three, and generic crossings are three-dimensional cones \cite{j.vonneumannBEHAVIOUREIGENVALUESADIABATIC2000}. I“It follows that, for generic non-time-reversal-invariant Hamiltonians depending on two parameters $(\phi_1,\phi_2)$ and additional parameters $\mathbf{p}$, conical intersections between bands occur on codimension-one submanifolds of the $\mathbf{p}$ parameter space. When $\mathbf{p}$ crosses such a submanifold transversely, the Chern number of one of the intersecting bands changes by 1, and the Chern number of the  other intersecting changes by $-1$~\cite{faureTopologicalChernIndices2000}.

The first major application of Chern invariants to physics was made in the theory of the quantum Hall effect \cite{klitzingNewMethodHighAccuracy1980}. Here the base manifold is the two-dimensional torus of Bloch momenta, and the Hall conductance of a band is equal to its Chern number multiplied by the quantum of conductance \cite{thoulessQuantizedHallConductance1982,avronHomotopyQuantizationCondensed1983, mahitokohmotoTopologicalInvariantQuantization1985}. Although nontrivial Chern classes can only arise when time reversal symmetry is broken, this can happen without external magnetic fields, in which case the Chern number is proportional to the \emph{anomalous} quantum Hall conductance \cite{haldaneModelQuantumHall1988}. A related physical phenomenon is quantized transport in multiply connected systems, where the electromotive force integrated over the threading of a unit magnetic flux through a hole generates transport of a Chern number of charge units \cite{r.b.laughlinQuantizedHallConductivity1981, d.j.thoulessQuantizationParticleTransport1983,avronAdiabaticQuantumTransport1988}. Later on, following the discovery of the topological aspects of the quantum spin Hall effect \cite{kaneZ_2TopologicalOrder2005,bernevigQuantumSpinHall2006}, it was realized that systems with quantized Hall conductance are one class of topological states of matter that are characterized by topological invariants classified by symmetries and the dimension of the base space \cite{hasanColloquiumTopologicalInsulators2010,qiTopologicalInsulatorsSuperconductors2011}.

{
The topological properties of complex systems with a large number of energy bands can be naturally studied by examining the distribution of topological invariants. The underlying supposition is that, under certain conditions, the distribution of topological invariants in such systems becomes universal and can be analyzed using synthetic random matrix models, where complexity is effectively replaced by randomness.

Originally, random matrix models were successfully employed to derive universal level statistics \cite{wigner1955,GUHR1998} and transport properties \cite{c.w.j.beenakkerRandommatrixTheoryQuantum1997} of complex quantum systems. The statistics of adiabatic curvature and Chern numbers, as functions of parameters, were later computed within random-matrix frameworks \cite{walkerUniversalFluctuationsChern1995,gatCorrelationsQuantumCurvature2021,swartzbergUniversalChernNumber2023}. Similarly, the statistics of winding numbers in one-dimensional random matrix models with chiral symmetry have been studied in \cite{braunWindingNumberStatistics2022,hahnWindingNumberStatistics2025}, \cite{hahnWindingNumberStatistics2023b}, and \cite{hahnWindingNumberStatistics2023} for the chiral unitary, symplectic, and orthogonal classes, respectively.

Many foundational models of topological states of matter are based on tight-binding Hamiltonians defined on periodic lattices \cite{thoulessQuantizedHallConductance1982,suSolitonsPolyacetylene1979,Haldane1988}. In the present work, we investigate the Chern numbers of energy bands in a quantum graph exhibiting a discrete two-dimensional square lattice structure, with a general four-port scatterer at each node.
Quantum graphs have proven highly effective in modeling complex physical systems, from Pauling’s early work on free-electron models for organic molecules \cite{Pauling1936} to studies of the Anderson transition in disordered wires \cite{shpiro1982}, mesoscopic quantum systems \cite{Texier2004}, and quantum chaos \cite{slim1997,KottosSmilansky_2003,GutkinSmilansky_2001}. Unlike tight-binding models, quantum graphs possess an infinite number of energy levels, making them well-suited for exploring universality. In particular, even relatively small quantum graphs exhibit universal spectral statistics that are well-described by appropriate ensembles of random matrices \cite{slim1997,Gnutzmann2008}.
Moreover, quantum graph models enable direct access to edge states and quantized transport—key features of topological matter that are not naturally captured within conventional random matrix models.

The Chern numbers of a quantum graph lattice with an external magnetic field were computed by Goldman and Gaspard \cite{Goldman2008}, who considered a special class of node scatterers that preserve time-reversal symmetry. In their work, nontrivial Chern numbers arise only in the presence of a nonzero magnetic field, leading to a Hofstadter-butterfly-like spectrum and a quantum Hall conductance governed by a Diophantine equation of the type introduced in \cite{thoulessQuantizedHallConductance1982}.
In contrast, we consider quantum graphs in which the node scatterers break time-reversal symmetry, allowing nontrivial Chern numbers even in the absence of a net external magnetic field. In this sense, the quantum graph can be viewed as a model of the anomalous Hall effect \cite{Haldane1988,n.a.sinitsynSemiclassicalTheoriesAnomalous2008}. In the present work, we focus on the topological properties of the system, leaving a detailed analysis of transport phenomena for future studies.

The paper is organized as follows. Owing to translational symmetry, Bloch’s theorem is applicable, with a single plaquette per unit cell. Using this observation, we show in Section \ref{sec:qg} that the associated fiber Hamiltonian of the lattice graph is equivalent to that of a “figure-eight” graph. This graph  consists of two equal-length looped quantum wires connected to the ports of the node scatterer, with the loops pierced by magnetic fluxes proportional to the two components of the Bloch momentum. These magnetic fluxes parametrize the torus base space of the energy band line bundles.

The calculation of the energy bands of the graph is reduced, in the standard manner, to solving a generalized eigenvalue problem for a $4 \times 4$ unitary matrix. Since the wire lengths are assumed equal, the bands are periodic, and the Chern numbers need only be computed for a fundamental set of four bands. In Section \ref{sec:tpqg}, we show that the only possible Chern numbers of a given band are 0—in which case all bands are topologically trivial—or $\pm 1$, in which case all bands have Chern numbers of unit magnitude that alternate in sign.

It follows that these three types of Chern number patterns are separated by hypersurfaces in the space of the node scatterer parameters, corresponding to intersections of energy bands.
 The main result of this paper, presented in Section \ref{sec:cross}, is an explicit trigonometric equation describing these intersection surfaces, which form codimension-one submanifolds of the group of $4 \times 4$ unitary matrices. Each energy-band intersection surface is characterized by an orientation that determines the change in the Chern number when the surface is crossed in a given direction. These results are validated through comparison with numerical calculations of the Chern numbers obtained via direct integration. In Section \ref{sec:realsca}, we provide explicit expressions for the  energy-band intersection surfaces in the special case where all matrix elements of the scatterer are real. In a sense clarified below, this case corresponds to the most extended region of topological nontriviality.

}

\section{The quantum-graph models}\label{sec:qg}

\subsection{The square lattice quantum graph}
\begin{figure}[h]

    \centering
 \includegraphics[width=9cm]{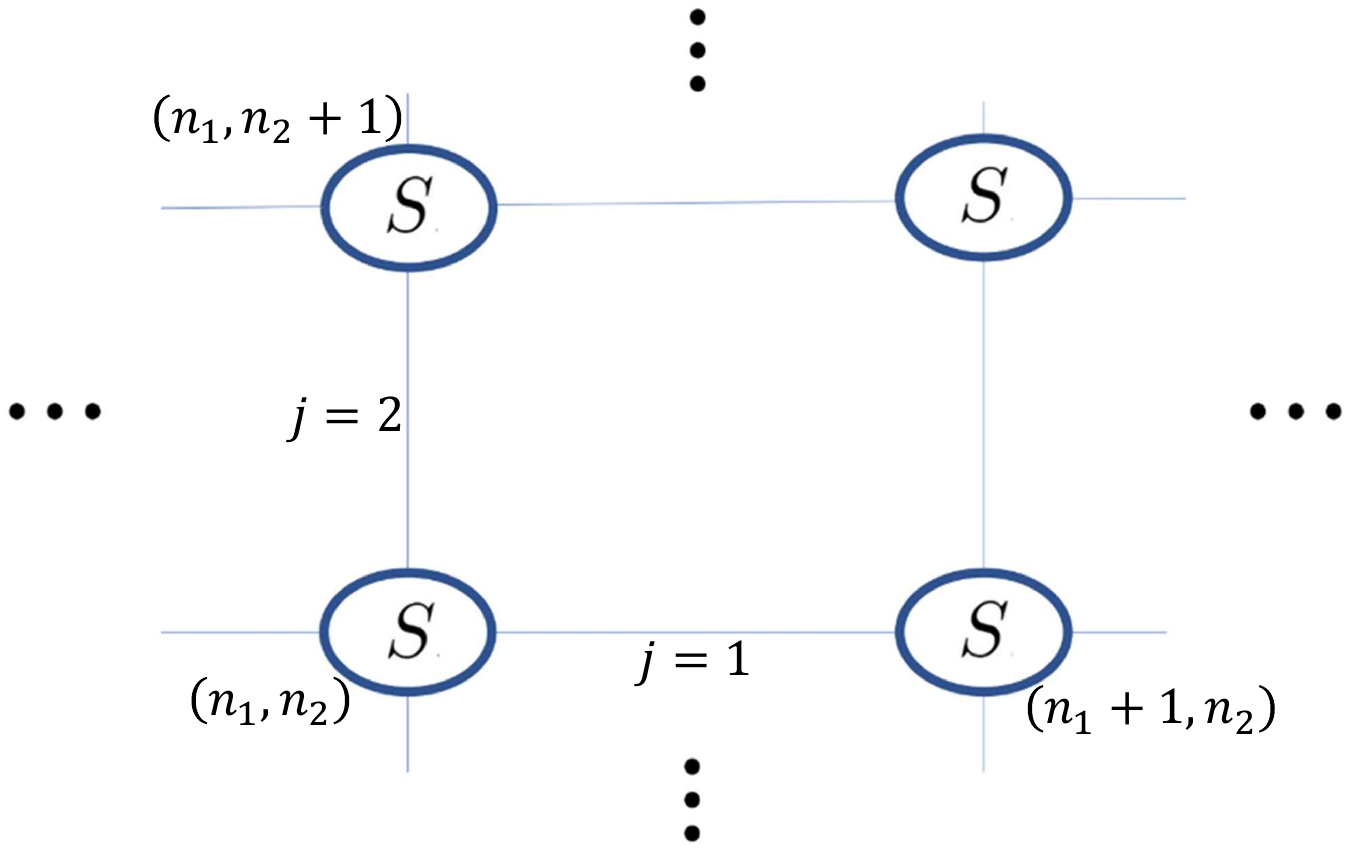}\;\;
\raisebox{1cm}{\includegraphics[width=6cm]{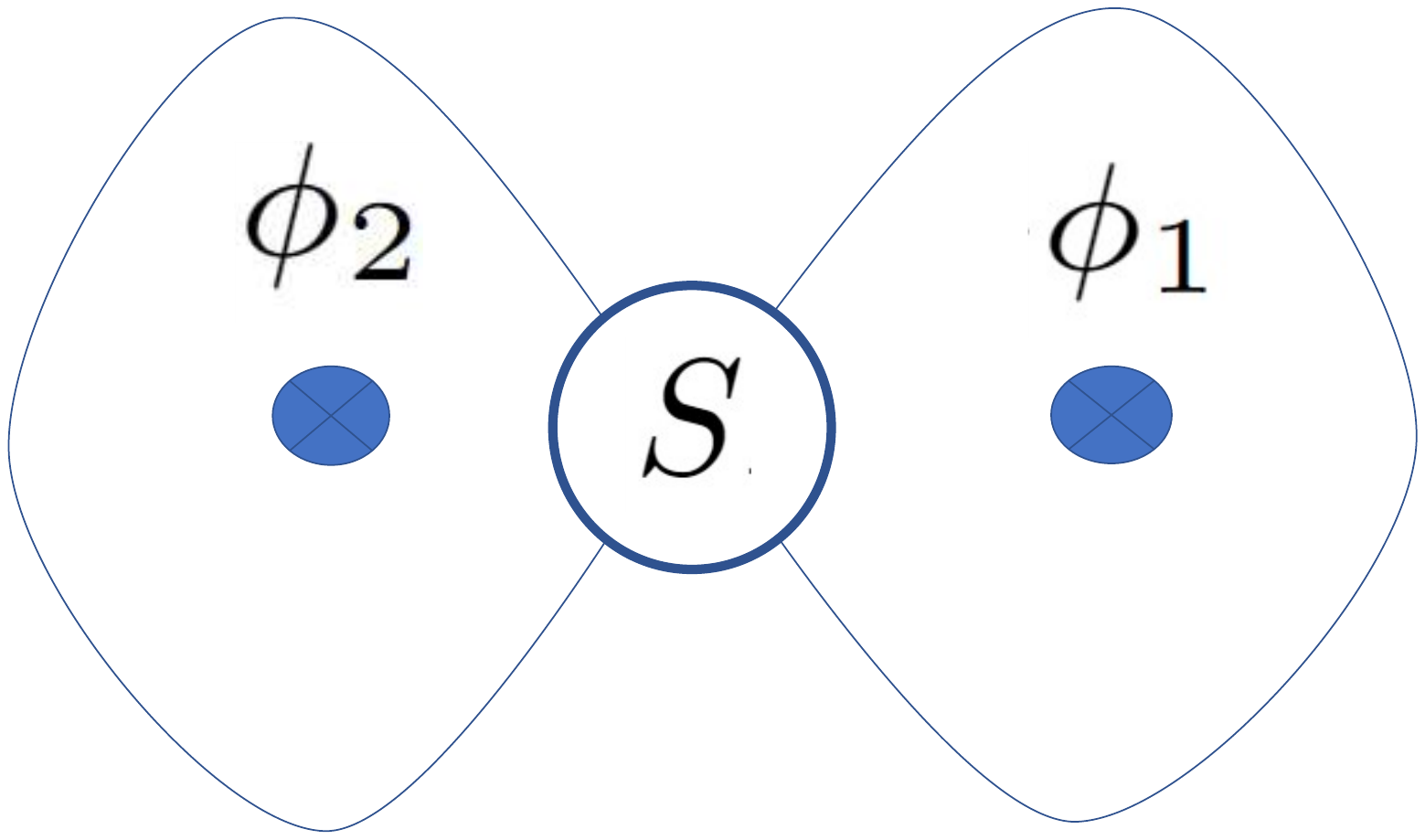}}
    \caption{Quantum graph models: Particles propagate freely on edges and are scattered at vertices by a scattering matrix $S$. Left: A periodic square lattice with nearest-neighbor edges and no magnetic fields. Right: A figure-eight graph with a single vertex and two edge loops pierced by Aharonov--Bohm magnetic fluxes $\phi_1,\,\phi_2$. The spectral problems defined by the two graphs are related by Bloch’s theorem: the lattice wave functions are classified by two quasimomenta, which correspond to the magnetic fluxes in the figure-eight graph.   \label{fig:8QG}}
\end{figure}

We study a quantum graph with a square-lattice structure. A quantum particle propagates freely along edges of equal length $L$ connecting nearest-neighbor vertices and is scattered by identical scatterers at the vertices, as shown in the left panel of figure \ref{fig:8QG}. We label the vertices of the graph by a pair of integers $n_1,n_2$, and the edges by a vertex label   and  an index $j=1,2$, representing the lattice direction.

The stationary states of the quantum graph problem are given by a set of wave functions  $\psi^{(j)}_{n_1,n_2}$ satisfying the Schrödinger equation
\begin{equation}\label{Sro Eq}
    -\frac{d^2\psi^{(j)}_{n_1,n_2}}{d{x}^2}={k}^2\psi^{(j)}_{n_1,n_2} , \qquad x\in [0,L]
\end{equation}
 on each edge of the graph,
where   $k= \sqrt{2mE/\hbar^2}$ is the wave number, and $m$ and $E$ are the mass and the energy of the particle,
respectively  \cite{gregoryberkolaikoElementaryIntroductionQuantum2017}. 
The solutions of these equations  are given by combinations of plane waves:
\begin{equation}\label{|Sol Eq}
   \psi^{(j)}_{n_1,n_2}=C^{(j)}_{n_1,n_2}e^{ikx}+\bar{C}^{(j)}_{n_1,n_2}e^{-ikx}. 
\end{equation}

The vertex scatterers are specified by a 4-by-4 unitary scattering matrix $S$ that connects the incoming amplitudes $\chi_{n_1,n_2}$ at vertex $n_1,n_2$ with the outgoing $\chi’_{n_1,n_2}$,
where 
\begin{equation}\label{Vect Eq}
   \chi_{n_1,n_2}=
   \begin{pmatrix}
   C^{(1)}_{
{ n_1-1,n_2}}e^{ikL}\\ \bar{C}^{(1)}_{n_1,n_2} \\ {C}^{(2)}_{n_1,n_2-1}e^{ikL}\\ \bar{C}^{(2)}_{n_1,n_2}
   \end{pmatrix}, \qquad 
   \chi'_{n_1,n_2} = 
\begin{pmatrix}
 \bar{C}^{(1)}_{
{n_1-1,n_2}}e^{-ikL}\\ {C}^{(1)}_{n_1,n_2}\\ \bar{C}^{(2)}_{n_1,n_2-1}e^{-ikL}\\ {C}^{(2)}_{n_1,n_2}
   \end{pmatrix}.
\end{equation}
The scattering relations $S\chi_{n_1,n_2}=\chi'_{n_1,n_2}$ can be viewed as a system of equations for the edge-wave amplitudes.
The discrete translation symmetry of the graph implies that we can use Bloch’s theorem to seek solutions of the system of the form $C_{n_1,n_2}^{(j)}=c^{(j)}e^{i({n_1\phi_1+n_2\phi_2})}$, $\bar C_{n_1,n_2}^{(j)}=\bar c^{(j)}e^{i({n_1\phi_1+n_2\phi_2})}$. For each set of values of the quasimomenta $\phi_1,\phi_2$ in the two spatial dimensions, the scattering equation becomes a generalized eigenvalue problem for $k$ with a discrete set of solutions, each of which delineates an energy band, see \cite{Band2013,Berkolaiko2018}.

Since the quasimomenta are $2\pi$-periodic, the bands of the lattice quantum graph that are separated by gaps from the rest of the spectrum define a smooth line bundle over the two-torus, in the same manner that bands of tight-binding lattice Hamiltonians do. In the following, we study the topology of these line bundles.

Quantum graph models: Particles propagate freely on edges and are scattered at
vertices by an identical scattering matrix $S$. Left: A periodic square lattice graph
with nearest-neighbor edges and no magnetic fields. The vertices are labeled by pairs of integers $(n_1,n_2)$, and the two lattice directions are labeled by $j=1,2$.
Right: A figure-eight graph with a single vertex and two edge loops pierced by
Aharonov-Bohm magnetic fluxes $\phi_1,\phi_2$. The directions $j=1,2$ are mapped
to the fluxes $\phi_1,\phi_2$, respectively. Thus, the two quasimomenta of the
lattice problem are represented as the two magnetic fluxes of the figure-eight
graph.

\subsection{The two-loop quantum graph} 
The lattice graph spectral problem for fixed values of $\phi_1,\phi_2$ is equivalent to that of a figure-eight graph, whose two length-$L$ loops are pierced by  magnetic  fluxes with corresponding phases  $\phi_1$ and  $ \phi_2$,  as shown in  figure \ref{fig:8QG}. 
This spectral problem can be formulated in the standard way, see, e.g., \cite{Gnutzmann2008}, as
\begin{equation}\label{eq:4spectral}
    \Lambda S\Psi=e^{-ikL}\Psi,
\end{equation}
where $\Lambda$  is the magnetic phase-shift matrix that is block-diagonal in a basis where the first (last) two components refer to the 1st (2nd) loop, respectively:
\begin{equation}\label{eq:Lambda}
    \Lambda=\begin{pmatrix}
        \Lambda_{1} & 0 \\ 
        0 &  \Lambda_{2} 
    \end{pmatrix},
    \qquad 
    \Lambda_{j}=\begin{pmatrix}
        0& e^{i\phi_j} \\ 
        e^{-i\phi_j}& 0  
    \end{pmatrix}.
\end{equation}
The four components of $\Psi$ are the forward- and backward-propagating amplitudes on the two loops of the figure-eight graph. They are linear combinations of the Bloch amplitudes $c^{(j)},\bar c^{(j)}$, and the energy bands of the lattice graph are reproduced when the fluxes are swept across the two-torus.
We note that the reduction of the nonmagnetic lattice graph to the magnetic figure-eight fiber graph also works when the lattice edge lengths in the two directions are not equal, but in this work we focus on graphs with equal-length edges, and choose $L=1$ without loss of generality.

The energy levels of the figure-eight graph are obtained from the secular equation
\begin{equation}
   \det( \Lambda S-e^{-ik} I)=0\ ,\label{SeqEq}
\end{equation}
where $I$ is the identity matrix. It follows that a solution $e^{-ik}$ of \eqref{SeqEq} is equal to one of the four eigenvalues $\sigma_j$, $j=1,\ldots 4$, of the unitary matrix $\Lambda S$, {implying} that the bands repeat periodically:
\begin{equation}
   k_{j,n}(\phi_1,\phi_2)=-\log\sigma_j(\phi_1,\phi_2)+2\pi n, \qquad n\in \mathbb{Z}\ .   \label{EnergyBands}
\end{equation}
The corresponding eigenvectors, and therefore the Chern numbers, are the same for all bands inside each of the four sequences.

\begin{figure}
    \centering
  \includegraphics[width=0.45\linewidth]{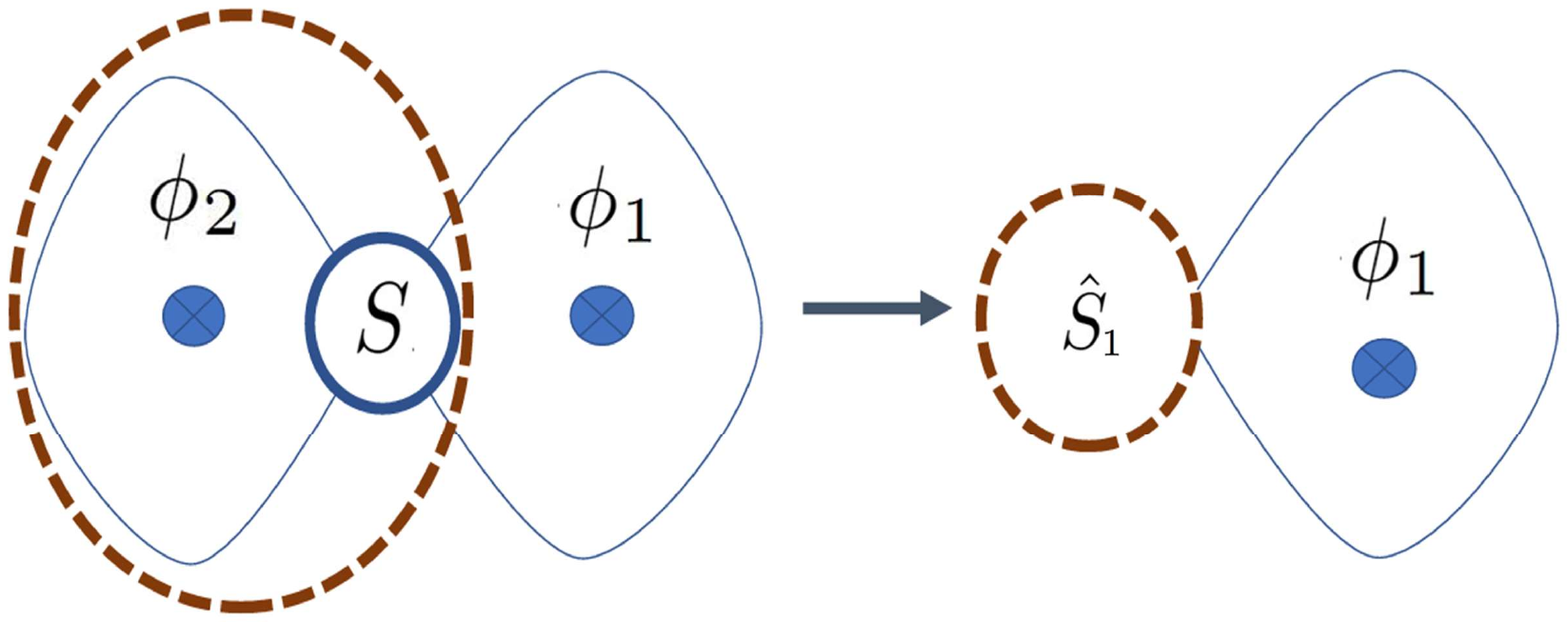} \hspace{0.5cm} \includegraphics[width=0.45\linewidth]{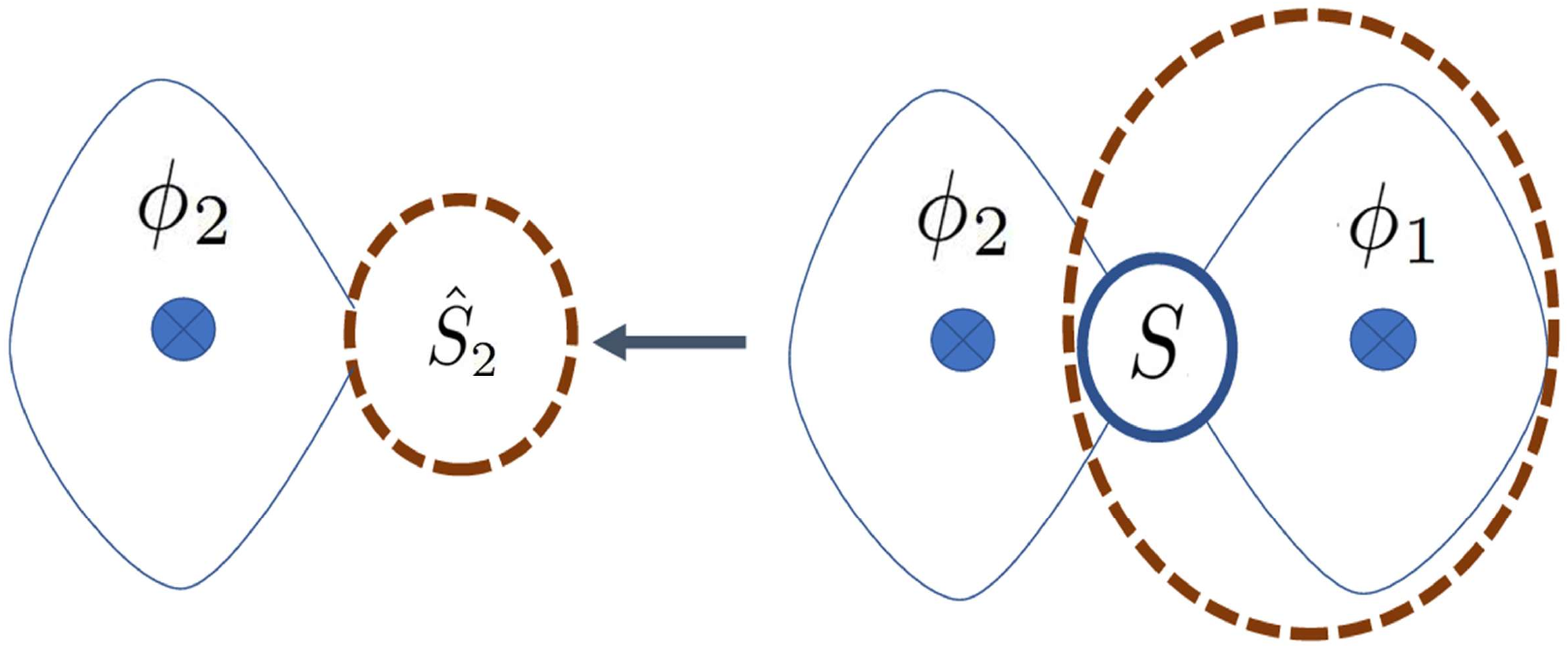}  
    \caption{The two-loop quantum graph in figure \ref{fig:8QG} can be viewed as a one-loop quantum graph pierced either by the flux $\phi_2$ (left) or by the flux  $\phi_1$ (right). The corresponding  $2\times 2$ vertex scattering matrices  $\hat{S}_1$,  $\hat{S}_2$  depend on  $k, \phi_1$, and $k, \phi_2$, respectively. }
    \label{fig:1l}
\end{figure}
\subsection{The one-loop quantum graph}
The spectrum   of the graphs  in figure \ref{fig:8QG} can be evaluated in yet another way by absorbing the loop pierced by the flux $\phi_1$, together with the  vertex,  into  a single scatterer. The reduced quantum graph consists of a single loop threaded by a flux of $\phi_2$, connecting the two ports of the scatterer, see figure \ref{fig:1l}, with a reduced 2-by-2 scattering matrix $\hat{S}_2$ that depends on the flux $\phi_1$ and $k$.  

To derive $\hat S_2$, we write the original $4$-by-$4$ scattering matrix and the amplitude vector in block form 
\begin{align}\label{Smatrix}
    S=\begin{pmatrix}
S_{11} & S_{12}\\
S_{21} & S_{22} 
\end{pmatrix}\ ,\qquad\Psi=\begin{pmatrix}\Psi_1\\\Psi_2\end{pmatrix}\ ,
\end{align}
with 2-by-2 matrix blocks and 2-vector column blocks.
The spectral equation \eqref{eq:4spectral} then implies that
\begin{equation}
\Psi_1=\frac{1}{e^{-ik}\Lambda_{1} -S_{11}}S_{12}\Psi_2\ ,
\end{equation}
and therefore
\begin{equation}\label{eq:2spectral2}
\Lambda_2\hat{S}_2(k)\Psi_2=e^{-ik}\Psi_2\ , \qquad\hat{S}_2(k)= S_{21} \frac{1}{e^{-ik}\Lambda_{1} -S_{11}}S_{12}+S_{22}\ .
\end{equation}
The same reduction applied to the second loop yields a scatterer-loop graph with a flux of $\phi_1$ and a reduced scattering matrix $\hat{S}_1$, whose spectral equation, related to \eqref{eq:2spectral2} by symmetry, is
\begin{equation}\label{eq:2spectral1}
\Lambda_1\hat{S}_1(k)\Psi_1=e^{-ik}\Psi_1\ , \qquad\hat{S}_1(k)= S_{11}+S_{12} \frac{1}{e^{-ik}\Lambda_{2}-S_{22}}S_{21}\ .
\end{equation}

It follows from the arguments leading to the reduced spectral problems (\ref{eq:2spectral2}--\ref{eq:2spectral1}) that they are equivalent to the original one \eqref{eq:4spectral} unless $e^{-ik}\Lambda_1-S_{11}$ or $e^{-ik}\Lambda_2-S_{22}$ is singular. As a result, the energy bands of the    one-loop graphs are obtained from the secular equations 
\begin{equation}\label{graphSpec}
   \det( \Lambda_{1}\hat S_1 (k) -e^{-ik} I)=\det( \Lambda_{2}\hat S_2 (k) -e^{-ik} I)=0\ ,  
\end{equation}
which are equivalent to equation~\eqref{SeqEq},  although in this case the $k$-dependent scattering matrices $\hat S_1$ and $\hat S_2$ are of reduced $2\times 2$ size. As we show below, this dimensional reduction simplifies the analysis of the \emph{topology} of the energy bands.

\section{Topological phases of the quantum graph}\label{sec:tpqg}
In this section, we demonstrate that the quantum graph models exhibit two distinct topological phases: a trivial phase, where the Chern numbers of all energy bands are 0, and a nontrivial phase, where the Chern numbers alternate between  $\pm1$  across the bands. This result is established in two steps: first, we show that the sum of the absolute values of the Chern numbers is four or less, and then that topological transitions occur as simultaneous intersections of two pairs of bands.

\subsection{The Chern numbers of energy bands}
\subsubsection{Chern numbers as degrees of maps}

Sticlet et al.\ \cite{Sticlet2012} showed that the Chern numbers of the bands of a 2-by-2 fiber Hamiltonian are equal to the Brouwer degree of a map from the base space to the two-sphere defined by the linear expansion of the Hamiltonian in terms of Pauli matrices. This result extends in a straightforward manner to our case, by writing
\begin{equation}\label{Hrepr}
    e^{ik}\Lambda_2\hat{S}_2=e^{i(I h_0+\Vec{\sigma}\cdot\Vec{h})}\ ,
\end{equation}  
where   $\Vec{\sigma}=(\sigma_1,\sigma_2,\sigma_3)$ stands for the vector of Pauli matrices  and   $\Vec{h}=(h_1,h_2,h_3)$ is a vector whose components are real-valued functions of the band phases $\phi_1,\phi_2$ and the spectral parameter $k$.

In the remainder of this subsection, we briefly review the spectral and topological properties of bands defined implicitly by imposing \eqref{eq:2spectral2}, or equivalently by requirinf that $I h_0+\Vec{\sigma}\cdot\Vec{h}$ has an eigenvalue equal to an integer multiple of $2\pi$, that is,
\begin{equation}\label{eq:1loopsproj}
h_0(k)+s\|\Vec{h}(k)\|\equiv 0 \pmod{2\pi}\ , \quad (s=\pm1)\ .
\end{equation}
For given $\phi_1,\phi_2$, four values of $e^{-ik}$ can be attained by solutions of \eqref{eq:1loopsproj}, namely the four eigenvalues of $\Lambda S$; see \eqref{eq:4spectral}. Labeling these four values $k_j$, $j=1,\ldots,4$, for example in increasing order for $-\pi<k_j\le\pi$ at $\phi_1=\phi_2=0$, and by continuity elsewhere, we obtain four energy bands $k_j(\phi_1,\phi_2)$, each associated with a sign choice $s_j$ in \eqref{eq:1loopsproj}.

Since $\|\Vec{h}\|>0$ away from points of degeneracy, we can define there the spectral projections
\begin{equation}\label{eq:hproj}
\mathcal{P}_j(\phi_1,\phi_2)=\frac{1}{2}(I+s_j\hat h(\phi_1,\phi_2,k_j)\cdot\Vec{\sigma})\ ,\qquad \hat h=\frac{\Vec{h}}{\|\Vec{h}\|}
\end{equation}
onto band $j$.

When the bands do not intersect,  $\|\Vec{h}\|>0$ everywhere on the torus $0\le\phi_1,\phi_2\le2\pi$, so that the four bands define maps $f_j=s_j \hat h(\phi_1,\phi_2,k_j)$  from the torus into the unit sphere, and by the arguments of \cite{Sticlet2012}, the Chern number of band $j$ is equal to the degree of $f_j$.
Recall \cite{leeIntroductionSmoothManifolds2013} that a point in the range of a map is regular if the Jacobian determinant of the map does not vanish at any of its preimages, and that the degree of a map is equal to the sum of the signs of its Jacobian determinant at all preimages of a regular point. Thus, for any pair $\hat h$ and $-\hat h$ of antipodal regular points 
\begin{equation}\label{ChernInt3}
C_j=\frac{s_j}{2}\sum\sign\bigl((\partial_{\phi_1}\Vec{h}_j\times\partial_{\phi_2}\Vec{h}_j)\cdot \vec h_j\bigr) ,
\end{equation}
where the sum is over all preimages $(\phi_1,\phi_2)$ of $\hat h_j\equiv\hat h(k_j)=\pm\hat h$.

\subsubsection{Calculation of the preimages}\label{sec:preim}
We next calculate the preimages of the north and south poles $\hat h_1=\hat h_2=0,\,\hat h_3=\pm1$, of the unit sphere. The condition $h_1=h_2=0$ is equivalent to the statement that the matrix $\hat S_2$ is purely off-diagonal. Since $\hat S_2$ is unitary, it is sufficient that the matrix element $\hat S_{2,11}=0$. Using the explicit representation \eqref{eq:2spectral2}, the latter condition can be expressed as
\begin{equation}\label{abcd}
    a e^{ik}+b e^{-ik}=c e^{i\phi_1}+d e^{-i\phi_1}\ ,
\end{equation}
where the coefficients $a,\,b,\,c,\,d$ depend only on the parameters of the $S$ matrix (explicit expressions are provided in section \ref{sec:diag} below). 

As $k$ is swept from $0$ to $2\pi$, the left-hand side of \eqref{abcd} describes an ellipse in the complex plane, centered at the origin with semi-major(minor) axis of length $||a|\pm|b||$ (respectively),  and a similar result holds for the right-hand side as $\phi_1$ is swept over the same range. If the poles are regular points of the map $\hat h$, the two ellipses either do not intersect at all, or intersect at four distinct  pairs $(k,\phi_{1})$. If the poles are singular points of the map, the ellipses are tangent at two pairs $(k,\phi_{1})$. In section \ref{sec:cross} below we derive the necessary and sufficient conditions on the coefficients of \eqref{abcd} for the ellipses to intersect, and derive expressions for the values of $k$ and $\phi_1$ at these intersections.

When the ellipses defined by the left- and right-hand sides of equation \eqref{abcd} do not intersect, there is no choice of fluxes for which the spectral projections are diagonal, so all the bands are trivial. When the ellipses do intersect, the off-diagonal elements of the matrix $\hat S_2$ evaluated at the intersection points are pure phases, so that the eigenvectors  $\Psi_2$ in \eqref{eq:2spectral2} are $(1,0)$ and $(0,1)$ for $e^{i\phi_2}=e^{-ik}\hat S_{2,21}^*$ and $e^{i\phi_2}=e^{ik}\hat S_{2,12}$, respectively. It follows that, in total, the sums in \eqref{ChernInt3} are over eight torus points. Since each of these preimages makes a $\pm1/2$ step in the Chern number of one of the four bands, we can bound the sum of the absolute values of the Chern numbers
\begin{equation}\label{eq:sumc}
\sum_{j=1}^4|C_j|\le4\ .
\end{equation}
In the next section we show that whenever there are topologically nontrivial bands, two of the bands have $C=1$ and the other two have $C=-1$.

\subsection{Band intersections and topological phases}\label{sec:top}
The graph spectral problem has a built-in symmetry that is most evident in the two-loop formulation \eqref{eq:4spectral}: the matrix $\Lambda$ of \eqref{eq:Lambda} obeys $\Lambda(\phi_1+\pi,\phi_2+\pi)=-\Lambda(\phi_1,\phi_2)$, implying that if $k_1,\ldots,\,k_4$ are the band wave numbers of the graph for flux values $(\phi_1,\phi_2)$, the the band wave numbers at $(\phi_1+\pi,\phi_2+\pi)$ are $k_1+\pi,\ldots,\,k_4+\pi$ (not necessarily in the same order, and modulo $2\pi$). In particular, if two bands intersect at $(\phi^*_1,\phi^*_2)$, it follows that $(\phi^*_1+\pi,\phi^*_2+\pi)$ is another intersection point. However, as we now show, the bands that intersect at the first point are distinct from those that intersect at the second point. 

\begin{figure}
    \centering    
\includegraphics[ height=5cm]{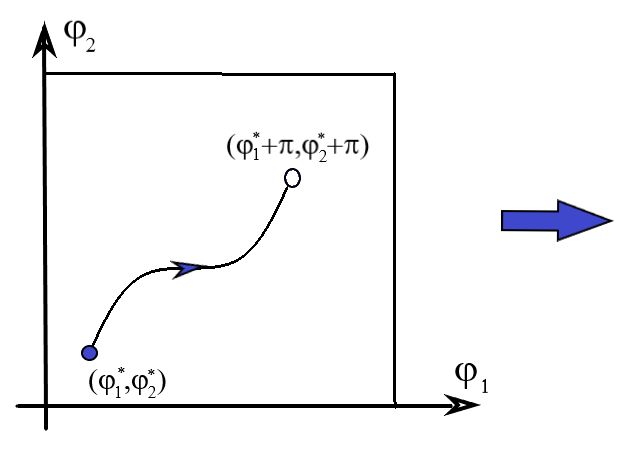}
\includegraphics[ height=5cm]{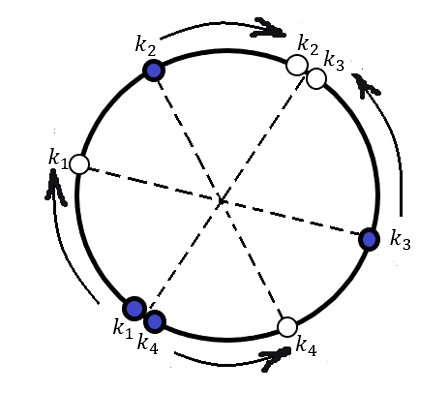}
\caption{The left panel  shows a continuous path on the torus of possible flux values, connecting  the two  degeneracy points $(\phi^*_1,\phi^*_2)$ and  $(\phi^*_1+\pi,\phi^*_2+\pi)$. The right panel shows the corresponding paths of the band wave numbers as the fluxes are swept along the path. If the pair of bands $\{k_1,\, k_4\}$ is degenerate at  $(\phi^*_1,\phi^*_2)$, then the pair $\{k_3,\, k_2\}$ is necessarily degenerate  at $(\phi^*_1+\pi,\phi^*_2+\pi)$. Thus, the topological transition from the trivial to the nontrivial phase occurs for all bands simultaneously.}
\label{fig_eigens}
\end{figure}

To this end, we order the bands $k_1,k_2,k_3,k_4$ in cyclically increasing order as above. Assume for concreteness that  $k_1(\phi^*_1,\phi^*_2)=k_4(\phi^*_1,\phi^*_2)$, i.e., the intersection is between the first and fourth bands. Consider a path $\mathcal{C}$ in the $(\phi_1,\phi_2)$ plane connecting the two degeneracy points $(\phi^*_1,\phi^*_2)$ and  $(\phi^*_1+\pi,\phi^*_2+\pi) $, such that there are no other degeneracy points along   $\mathcal{C}$, see figure \ref{fig_eigens}, left panel.  Note that since $\det\Lambda=1$ independently of $\phi_1,\phi_2$, the sum $\sum k_j$ is constant. The spectral symmetry of the graph bands dictates that the set of band wave numbers should reach values shifted by $\pi$ when the fluxes are varied along $\mathcal{C}$. The only way to achieve this continuously while maintaining $\sum k_j$ is by mapping 
\begin{equation}
k_1\to k_3-\pi\ ,\qquad k_2\to k_4-\pi\ ,\qquad k_3\to k_1+\pi\ ,\qquad k_4\to k_2+\pi\ ,
\end{equation}
see figure \ref{fig_eigens}, right panel. It follows that the intersection at $(\phi^*_1+\pi,\phi^*_2+\pi)$ is between the second and third bands.  

If the parameters of the scattering matrix $S$ are varied from the trivial to the nontrivial phase so that $C_4$ increases by one at the band-intersection point, then the Chern numbers in the nontrivial phase become $-C_1 = C_2 = -C_3 = C_4 = 1$. Conversely, if the parameters are varied so that $C_4$ decreases by one at the band-intersection point, Chern numbers with opposite signs are obtained for each band. In both cases, the nontrivial phase is characterized by alternating $\pm 1$ Chern numbers.
If the parameters of $S$ are varied  further from such a phase through another band-intersection point, the resulting Chern numbers could, in principle, either all become zero or take values $\pm 2$. The latter possibility, however, is excluded, as it would violate the bound \eqref{eq:sumc} on the sum of absolute Chern numbers.

 In conclusion, the quantum graph model has two possible topological phases: a trivial phase with zero Chern numbers for all bands and a nontrivial phase characterized by an alternating $\pm 1$ pattern of Chern numbers across the four bands. If a path in the parameter space of the matrix $S$ passes through a point of spectral degeneracy, the gaps between adjacent bands close simultaneously, and the graph undergoes a transition from  the trivial to nontrivial phase, or vice versa.

\section{Topological transitions of the quantum graph}\label{sec:cross}

\subsection{Band intersections}
In this section, we establish a necessary and sufficient condition on the scattering  matrix $S$ for a transition to occur between the trivial and nontrivial phases of the quantum graph. Topological transitions occur when the gaps between the bands close and the spectrum of the matrix $\Lambda S$ has  a degeneracy for some values  $\phi_{1}$, $\phi_{2}$ of magnetic flux, and therefore also for $\phi_1+\pi,\phi_2+\pi$ as explained above in section \ref{sec:top}.

According to the Wigner--von Neumann theorem \cite{von1929no}, the manifold of degenerate unitary matrices has codimension-three in the manifold of all unitaries of a given size. Since the energy bands depend on two flux parameters, the manifold of scattering matrices for which a topological transition occurs has codimension-one. It is therefore convenient to choose one scattering parameter and express the band-intersection condition in terms of it. We choose for this purpose the scattering phase 
\begin{equation}
\gamma=\tfrac{1}{4}(\arg\det S_{11}-\arg\det S_{22})\ ,
\end{equation}
where $S_{11}$, $S_{22}$ are the two-by-two blocks defined in \eqref{Smatrix} above, and consider the one-parameter family of scattering matrices with two-by-two block representation
\begin{equation}\label{eq:sgamma}
S=\begin{pmatrix}
e^{i\gamma}I&0\\0&e^{-i\gamma}I
\end{pmatrix} S_0
\end{equation}
for some fixed $S_0$ for which $\gamma=0$.

We tackle the gap-closing problem in terms of the equivalent one-loop scattering matrix $\hat S_2$. At a generic intersection, the eigenspaces of the three spectral equivalent problems (\ref{eq:4spectral}), (\ref{eq:2spectral2}), and (\ref{eq:2spectral1}) are two-dimensional, so the matrices $ \Lambda_{1}\hat S_1, \Lambda_{2}\hat S_2$ are scalar multiples of the identity matrix,
\begin{align}\label{eq:2deg1}
&\Lambda_{1}\hat S_1=e^{-ik}I\ ,\\& \label{eq:2deg2}\Lambda_{2}\hat S_2=e^{-ik}I\ .
\end{align}
We derive the equation for $\gamma$ from \eqref{eq:2deg2} in three steps: First, we express $\phi_1$ at the band intersection points in terms of $\gamma$ and the spectral parameter $k$ (section \ref{sec:diag}). We then eliminate $\phi_1$ to obtain a relation between $\gamma$ and $k$ (section \ref{sec:gammak}). Finally, we eliminate $\phi_2$ between two further expressions to obtain a second relation between $\gamma$ and $k$, from which the equation for $\gamma$ is reached by eliminating $k$ (section \ref{sec:gammax}). In section \ref{sec:psibar} we obtain an independent necessary condition for spectral degeneracy by partially imposing both \eqref{eq:2deg1} and \eqref{eq:2deg2}.

\subsubsection{The flux values at points of degeneracy}\label{sec:diag}
Each of equations \eqref{eq:2deg1}, \eqref{eq:2deg2} is a necessary and sufficient condition for the degeneracy of the spectrum of the graph matrix $\Lambda S$. Since the matrices $\Lambda_j\hat S_j$ are unitary, these matrix equations impose four real conditions, which we solve for the real variables $\gamma,k,\phi_1,\phi_2$. In this subsection and the following two, we solve equation \eqref{eq:2deg2}, and then reconsider \eqref{eq:2deg1} in subsection \ref{sec:psibar}. 

It is convenient to start with the diagonal part of \eqref{eq:2deg2}, which is satisfied if and only if $\hat S_2$ is purely off-diagonal, since this condition does not involve $\phi_2$. A straightforward but lengthy calculation implies $\hat S_{2,11}=\hat S_{2,22}=0$ if and only if
\begin{equation}\label{eq:abcd1}
   a e^{i(k+\gamma)}+b e^{-i(k+\gamma)}=c e^{i\phi_1}+d e^{-i\phi_1}\ ,%\\\label{eq:abcd2}
%   & a_2 e^{i(k-\gamma)}+b_2 e^{-i(k-\gamma)}=c_2 e^{i\phi_2}+d_2 e^{-i\phi_2}\ ;
\end{equation}
where the coefficients depend polynomially on the elements of $S$:
\begin{align}
a&=S_{31}(S_{12}S_{23}-S_{22}S_{13})+S_{32}(S_{21}S_{13}-S_{11}S_{23})+S_{33}(S_{11}S_{22}-S_{12}S_{21})\ ,\\
b&=-S_{33}\ ,\\
c&=S_{31}S_{23}-S_{33}S_{21}\ ,\\
d&=S_{32}S_{13}-S_{33}S_{12}\ .
%a_2&=S_{13}(S_{34}S_{41}-S_{44}S_{31})+S_{14}(S_{43}S_{31}-S_{33}S_{41})+S_{11}(S_{33}S_{44}-S_{34}S_{43})\\
%b_2&=S_{11}\\
%c_2&=S_{13}S_{41}-S_{11}S_{43}\\
%d_2&=S_{14}S_{31}-S_{11}S_{34}
\end{align}
Next we note that equation \eqref{eq:abcd1} and its complex conjugates are independent over the reals, allowing us to express $e^{i\phi_1}$ as
\begin{equation}\label{eq:ABC1}
Ae^{i(k+\gamma)}+B e^{-i(k+\gamma)}=Ce^{i\phi_1}\ ,%\\\label{eq:ABC2}
%&A_2e^{i(k-\gamma)}+B_2 e^{-i(k-\gamma)}=C_2e^{i\phi_2}\ ,
\end{equation}
with
\begin{align}
A&=ac^*-b^*d\label{eq:A1}\ ,\\
B&=bc^*-a^*d\ ,\\
C&=|c|^2-|d|^2\ .\label{eq:C1}
\end{align}

\subsubsection{The wave number at points of degeneracy}\label{sec:gammak}
It follows from equation \eqref{eq:ABC1} that
\begin{equation}\label{eq:ABCk}
AB^*e^{2i(k+\gamma)}+A^*Be^{-2i(k+\gamma)}={C}^2-|A|^2-|B|^2\ ;
\end{equation}
this equation implies the solvability condition
\begin{equation}\label{eq:solv}
2|A||B|\ge |C^2-|A|^2-|B|^2|\ ,
\end{equation}
which is necessary and sufficient for the ellipses described by the left- and right-hand sides of \eqref{eq:abcd1} to intersect (as explained in section \ref{sec:preim} above).

The strict inequality form of \eqref{eq:solv} is equivalent to the three inequalities
\begin{equation}\label{eq:triabc}
|C|<|A|+|B|\ ,\qquad |B|<|C|+|A|\ ,\qquad |A|<|B|+|C|\ ,
\end{equation}
which are necessary and sufficient conditions for the existence of a triangle with side lengths $|A|,|B|,|C|$.

If we write \eqref{eq:ABCk} as
\begin{equation}\label{eq:k+gamma}
e^{2i(k+\gamma)-i\delta}+e^{i\delta-2i(k+\gamma)}=g\ ,
%g\cos(2(k+\gamma)-\delta)=1\ ,
\end{equation}
with 
\begin{equation}\label{eq:gd1}
ge^{i\delta}= \frac{C^2-|A|^2-|B|^2}{AB^*}\ ,\qquad g>0\ ,
\end{equation} 
then \eqref{eq:solv} is equivalent to $g\le2$, in which case, the possible wave numbers at the points of degeneracy are
\begin{equation}\label{eq:kpsi}
k=-\gamma+\frac{\delta\pm\psi}{2}+n\pi 
\end{equation}
with an integer $n$, where
\begin{equation}\label{eq:psi1}
\psi=\arccos\Bigl(\frac{g}{2}\Bigr)\ .
\end{equation}

\begin{figure}[t]\centering
    \includegraphics[scale=0.45]{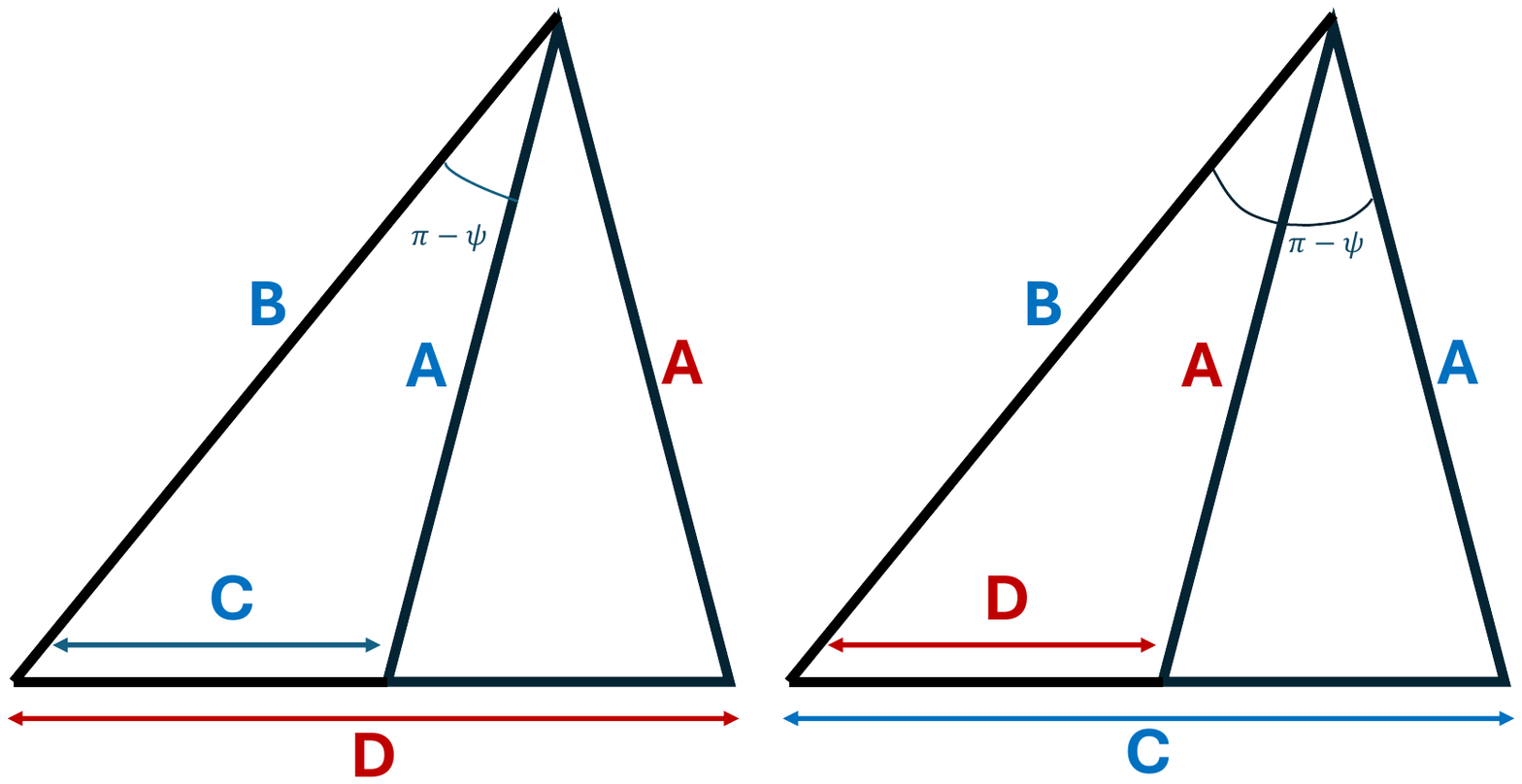}
    \caption{Geometric interpretation of the phase $\psi$ as an angle in a triangle with side lengths $|A|$, $|B|$, and $|C|$. A degeneracy in the graph spectrum is possible only when the triangle inequality (\ref{eq:triabc}) holds. In this case a similar triangle inequality between $|A|$, $|B|$, and $|D|$ holds as well.  The left and right panels depict the cases $|B|^2 > |C|^2 +|A|^2$ and $|B|^2 < |C|^2 +|A|^2$,  respectively.}
\label{figtriangle}
\end{figure}

Geometrically,  $\pi-\psi$ is the angle opposite the side of length $|C|$ in the triangle with side lengths $|A|^2+|B|^2>C^2$, see figure~\ref{figtriangle}. 
The analogous argument that starts by eliminating $k$ between \eqref{eq:abcd1} and its conjugate leads to an equivalent inequality for a triangle with sides $|A|$, $|B|$, and $|D|$, where $D=|a|^2-|b|^2$, and an expression for $\phi_1$ analogous to \eqref{eq:kpsi}, see appendix \ref{app:chi}.

\subsubsection{Scattering phase for band intersection}\label{sec:gammax}
The arguments of the previous subsections followed from the off-diagonal elements of the matrix equation \eqref{eq:2deg2} (for $\Lambda_2\hat S_2$), and allowed us to express $k$ and $\phi_1$ at degeneracy points in terms of the elements of the matrix $S$ and the scattering phase $\gamma$. We proceed to impose the diagonal part of \eqref{eq:2deg2} that determines $\phi_2$ at the point of intersection, and the $\gamma$ values for which the bands intersect.

Additional lengthy but straightforward algebra implies that the first and second diagonal elements of \eqref{eq:2deg2} are equivalent, respectively, to
\begin{align}
&q_1e^{i(k+\gamma)}+p_1e^{-i(k+\gamma)}-r_1e^{i\phi_1}-s_1e^{-i\phi_1}\nonumber\\&\qquad \qquad \qquad =(u e^{2i\gamma}+ve^{i(-k+\gamma+\phi_1)}+we^{i(-k+\gamma-\phi_1)}-e^{-2ik})e^{-i\phi_2}\ ,\label{eq:qprs1}\\
&q_2e^{i(k+\gamma)}+p_2e^{-i(k+\gamma)}-r_2e^{i\phi_1}-s_2e^{-i\phi_1} \nonumber\\&\qquad \qquad \qquad=(u e^{2i\gamma}+ve^{i(-k+\gamma+\phi_1)}+we^{i(-k+\gamma-\phi_1)}-e^{-2ik})e^{i\phi_2}\ ,\label{eq:qprs2}
\end{align}
with
\begin{align}
q_1&=S_{41}(S_{12}S_{23}-S_{22}S_{13})+S_{42}(S_{21}S_{13}-S_{11}S_{23})+S_{43}(S_{11}S_{22}-S_{12}S_{21})\\
p_1&=-S_{43}\\
r_1&=S_{41}S_{23}-S_{43}S_{21}\\
s_1&=S_{42}S_{13}-S_{43}S_{12}\\
q_2&=S_{31}(S_{12}S_{24}-S_{22}S_{14})+S_{32}(S_{21}S_{14}-S_{11}S_{24})+S_{34}(S_{11}S_{22}-S_{12}S_{21})\\
p_2&=-S_{34}\\
r_2&=S_{31}S_{24}-S_{34}S_{21}\\
s_2&=S_{32}S_{14}-S_{34}S_{12}\\
u&=S_{11}S_{22}-S_{12}S_{21}\\
v&=S_{21}\\
w&=S_{12}\ .
\end{align}

Substituting $e^{\pm i\phi_1}$ from \eqref{eq:ABC1} and its complex conjugate in \eqref{eq:qprs1}--\eqref{eq:qprs2} gives
\begin{align}
&Q_1e^{i(k+\gamma)}+P_1e^{-i(k+\gamma)}=(U e^{2i\gamma}+Ye^{-2ik})e^{-i\phi_2}\ ,\label{eq:QP1}\\
&Q_2e^{i(k+\gamma)}+P_2e^{-i(k+\gamma)}=(U e^{2i\gamma}+Ye^{-2ik})e^{i\phi_2}\ ,\label{eq:QP2}
\end{align}
with
\begin{align}
Q_n&=q_nC-r_nA-s_nB^*\ ,\\
P_n&=p_nC-r_nB-s_nA^*\ ,\\
U&=uC+vA+wB^*\ ,\\
Y&=-C+vB+wA^*\ ;
\end{align}
eliminating $e^{i\phi_2}$ between \eqref{eq:QP1} and \eqref{eq:QP2} and using \eqref{eq:kpsi} we obtain
\begin{equation}\label{eq:QPUV}
e^{i(4\gamma-(\delta\pm\psi))}=\frac{Q_1Q_2e^{i(\delta\pm\psi)}+Q_1P_2+P_1Q_2+P_1P_2e^{-i(\delta\pm\psi)}}{U^2e^{i(\delta\pm\psi)}+2UY +Y^2e^{-i(\delta\pm\psi)}}\ .
\end{equation}
Since $\gamma$ is defined modulo $\pi$, equation \eqref{eq:QPUV} has four inequivalent solutions
\begin{equation}\label{eq:gamma4}
\gamma=\frac{\chi_\pm+\delta\pm\psi}{4}\,,\quad \frac{\chi_\pm+ \delta\pm\psi}{4}+\frac{\pi}{2}\ ,
\end{equation}
with
\begin{equation}\begin{split}
&\chi_\pm=\arg\bigl(Q_1Q_2e^{i(\delta\pm\psi)}+Q_1P_2+P_1Q_2+P_1P_2e^{-i(\delta\pm\psi)}\bigr)\\&\hspace{4cm}-\arg\bigl(U^2e^{i(\delta\pm\psi)}+2UY +Y^2e^{-i(\delta\pm\psi)}\bigr)\end{split}
\end{equation}

Note that the ambiguous signs in \eqref{eq:gamma4} and \eqref{eq:kpsi} must be chosen upper or lower consistently. Thus, as discussed in section \ref{sec:top}, for each $\gamma$ for which the bands cross, there are two points of intersection with both flux values shifted by $\pi$, obtained by choosing $n=0$ and $n=1$ (or any other combination of even and odd values of $n$) in \eqref{eq:kpsi}. The fluxes at the intersections points are then given by \eqref{eq:ABC1} and, for example, \eqref{eq:qprs1}. There is also a simple relation between the intersection points for $\gamma$ values that differ by $\pi/2$: they occur at points where exactly one of the fluxes is shifted by $\pi$ for $k$ values that are shifted by $\pi/2$.

\subsubsection{A necessary condition for band intersection}\label{sec:psibar}
Recall that equation~\eqref{eq:abcd1} follows from the diagonal part of the two-by-two matrix  equation \eqref{eq:2deg2} and yields the relation \eqref{eq:kpsi} between $k$ and $\gamma$. Analogously, we can derive a second relation between these variables from the diagonal part  of the dual  matrix equation \eqref{eq:2deg1},
\begin{equation}\label{eq:abcd2}
   \bar a e^{i(k-\gamma)}+ \bar b e^{-i(k-\gamma)}= \bar c e^{i\phi_2}+ \bar d e^{-i\phi_2}\ ,
\end{equation}
with 
\begin{align}
\bar a&=S_{13}(S_{34}S_{41}-S_{44}S_{31})+S_{14}(S_{43}S_{31}-S_{33}S_{41})+S_{11}(S_{33}S_{44}-S_{34}S_{43})\\
\bar b&=S_{11}\\
\bar c&=S_{13}S_{41}-S_{11}S_{43}\\
\bar d&=S_{14}S_{31}-S_{11}S_{34}\ .
\end{align}

Following the steps that lead to \eqref{eq:kpsi} we define $\bar A,\,\bar B,\,\bar C,\,\bar g,\,\bar\delta,$ and $\bar\psi$ by the analogs of \eqref{eq:A1}--\eqref{eq:C1}, \eqref{eq:gd1}, and \eqref{eq:psi1} with each variable replaced by its barred counterpart, to obtain
\begin{equation}\label{eq:kpsi2}
k=\gamma+\frac{\bar\delta\pm \bar\psi}{2}+n\pi \ .
\end{equation}
Eliminating $k$ between \eqref{eq:kpsi} and \eqref{eq:kpsi2} gives
\begin{equation}\label{eq:gamma8}
\gamma=\frac{\delta-\bar\delta\pm\psi\mp\bar\psi}{4}+\frac{n\pi}{2}\ .
\end{equation}
Since the two sign choices in this expression are independent, it yields eight inequivalent values modulo $\pi$, compared with four choices in \eqref{eq:gamma4}. Thus, although the diagonal elements of \eqref{eq:2deg1} and \eqref{eq:2deg2} form a system of four real scalar equations for four real variables, some of the discrete solutions are not actual crossings. This system is therefore a necessary but not sufficient condition for band intersections.

The actual crossing values of $\gamma$ obtained from \eqref{eq:gamma8} can be identified by noting (see equations \eqref{eq:gd1} and \eqref{eq:psi1}) that $\psi$ and $\bar\psi$ change sign when $C$ and $\bar C$, respectively, change sign. Using this observation, we find the four inequivalent values of $\gamma$:
\begin{equation}\label{gammaSol}
    \left\{\frac{\delta-\bar\delta\pm(\psi\sign C-\bar\psi\sign \bar C)}{4}\,,\quad \frac{\delta-\bar\delta\pm(\psi\sign C-\bar\psi\sign \bar C)}{4}+\frac{\pi}{2} \right\}\ ,
\end{equation}
which are equal to those of \eqref{eq:gamma4}.

\begin{figure}
 \begin{subfigure}{0.45\textwidth}
        \centering
        \includegraphics[width=\textwidth]{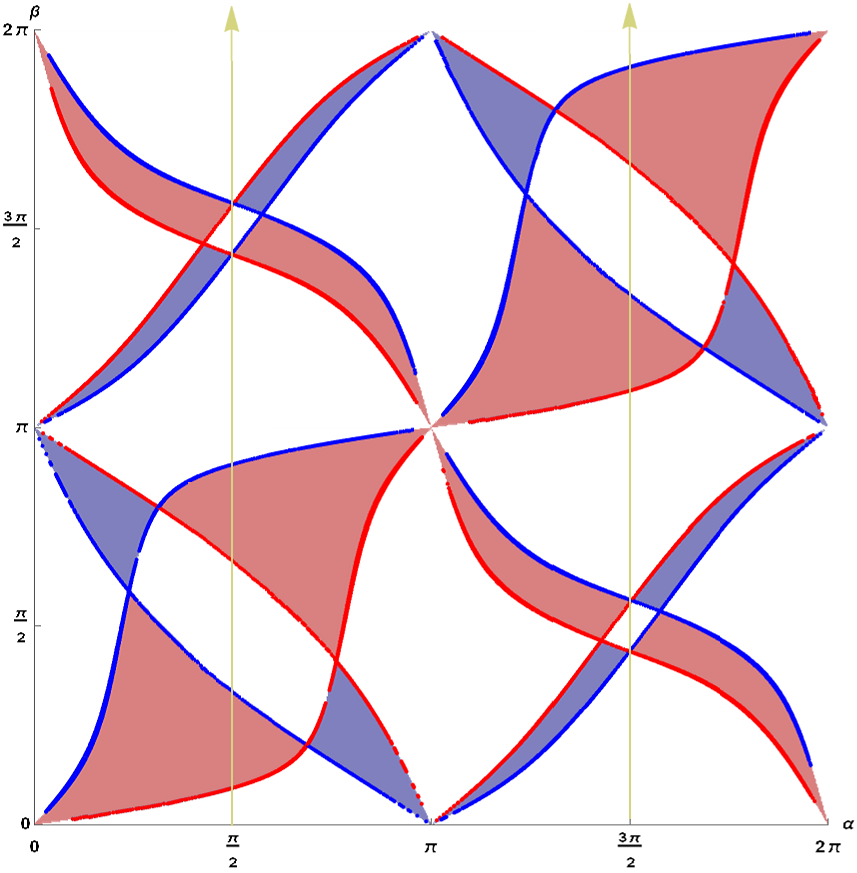}
    \end{subfigure}
\hspace{1.5cm}
    \begin{subfigure}{0.45\textwidth}
        \centering
        \includegraphics[width=\textwidth]{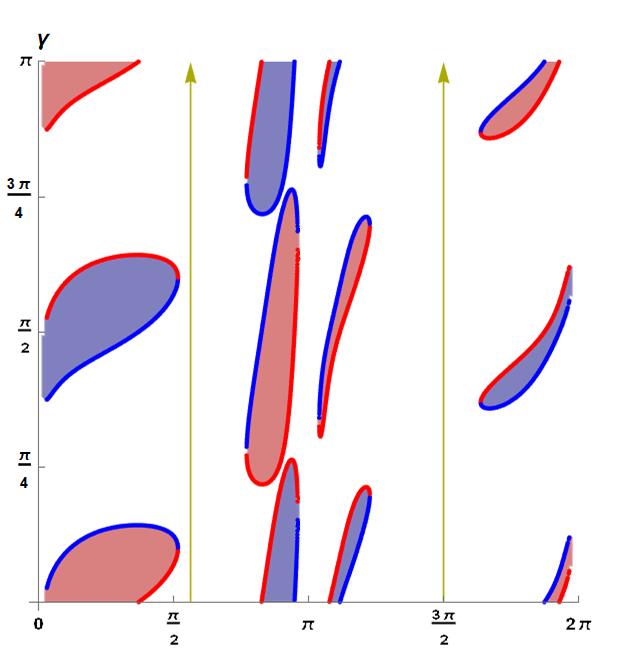}
    \end{subfigure}
    \caption{
Topological phase diagrams of the Chern numbers for one energy band  of the two-loop quantum graph with a 4-by-4 scattering matrix parametrized as described in appendix \ref{app:para}. In this figure we fix the parameters $\theta_1 = \pi/4$, $\theta_2 = \pi/3$, $\eta_1 = \pi/2$, $\eta_2 = \pi/6$, $\beta = 0.3$, $\nu_1 = 0.6$, and $\nu_2 = \mu_1=\mu_2=0$. The left  and the right  panels show  the Chern number  of one of the bands in the $\alpha$--$\beta$  plane for $\gamma=0.5$ and in the $\gamma$--$\alpha$  plane for $\beta=0.3$, respectively.  The Chern number is $-1$ in blue-shaded regions, $+1$ in red-shaded regions, and $0$ in unshaded regions. The Chern number was calculated numerically  using the algorithm detailed in appendix \ref{app:numerics}. 
The red and blue curves represent the band intersection contours obtained from (\ref{eq:gamma4}).   As detailed in Section \ref{sec:sign}, we calculated the sign of the determinant  in equation (\ref{eq:crossign}) with respect to  the parameter $\gamma$ on the right panel and parameter $\beta$ on the left panel. Red lines correspond to a positive determinant, meaning that  the Chern number increases  by $1$ after crossing lines  into the direction of yellow arrows, while blue lines represent negative determinant, corresponding to a decrease of the Chern number by $1$ along the same direction. }
   \label{PD2}
\end{figure}

\subsection{The orientation of the topological transitions}\label{sec:sign}
In the previous subsection we derived a necessary and sufficient condition for the crossing of the energy bands of the quantum graph, that is satisfied for a codimension-one family of scattering matrices $S$. When $S$ is varied continuously across the intersection manifold, the Chern number of the crossing bands jumps, so that the band topologies transition from trivial to nontrivial or vice versa. We set the sign of the transition according to the jump in the Chern number of, say, the upper of the two intersecting bands. We next derive an expression for the sign of the transition that allows us to draw the topological phase diagram of the model.

For this purpose, assume that $(\phi_{1,\times},\phi_{2,\times})$ is a degeneracy point of $\Lambda_\times S_{\times}$ with a double eigenvalue $e^{-ik_{\times}}$ so that $\Lambda_{2,\times}\hat S_{2,\times}=e^{-ik_{\times}}I$, and write $e^{ik_{\times}}\Lambda_{2}\hat S_{2}=\hat S_{2,\times}e^{i(h_0+\vec\sigma\cdot\vec h)}$ in terms of Pauli matrices as in \eqref{Hrepr}. Since the band-crossing manifold has codimension one, we can simplify the description by explicitly displaying the dependence of $S$ on a single parameter, denoted here as $\phi_3$ (corresponding to the phase $\gamma$ in Section~\ref{sec:cross}).
Viewing $h_0$ and $\vec{h}$ as functions of $\phi = (\phi_1, \phi_2, \phi_3)$ and $k$, the crossing conditions become $S(\phi_{3,\times}) = S_\times$, or equivalently $h_0(\phi_\times; k_\times)=\vec h(\phi_\times; k_\times)= 0$.

For $\phi_j$ close to $\phi_{j,\times}$, two of the solutions of \eqref{eq:1loopsproj}, $k_\pm$, with opposite choices of $s$, are close to $k_\times$, and we can approximate
\begin{equation}\label{eq:k+-}
\mathring{k}_\pm\frac{\partial h_0}{\partial k}+\mathring{\phi}\cdot\frac{\partial h_0}{\partial\phi}=\pm{\Bigl|\mathring{k}_\pm\frac{\partial\vec h}{\partial k}+\mathring{\phi}\cdot\frac{\partial\vec h}{\partial\phi}\Bigr|}\ ,
\end{equation}
where $\mathring{k}_\pm=k_\pm-k_\times$, $\mathring{\phi}=\phi-\phi_\times$, and the derivatives are evaluated at the crossing point.
Letting $h_{j,\pm}(\phi)=h_j(\phi;k_\pm)$, we now use \eqref{ChernInt3} to calculate the contributions $C_\text{loc}$ to the Chern numbers from the neighborhood of the crossing point, 
\begin{equation}\label{eq:clocpm}
C_\text{loc}=\pm\frac{1}{2}\sign\Bigl(\frac{\partial\Vec{h}_\pm}{\partial{\phi_1}}\times\frac{\partial\Vec{h}_\pm}{\partial{\phi_2}}\cdot \vec h_\pm\Bigr)\ ,
\end{equation}
where the expression is evaluated for any $\phi$ close to $\phi_\times$. 

To proceed, we calculate $\vec h_\pm$ for $\mathring{\phi}=\eta\phi_k$, $\eta\ll1$, where $\phi_k=(\partial\vec h/\partial\phi)^{-1}(\partial\vec h/\partial k)$, obtaining
\begin{equation}\label{eq:k+-eta}
\mathring{k}_\pm\frac{\partial h_0}{\partial k}+\eta\phi_k\cdot\frac{\partial h_0}{\partial\phi}=\pm|\mathring{k}_\pm+\eta|\Bigl|\frac{\partial\vec h}{\partial k}\Bigr|\ .
\end{equation}
Assuming $|{\partial h_0}/{\partial k}|>|{\partial\vec h}/{\partial k}|$, 
\begin{equation}\label{eq:h+-eta}
\vec h_\pm=(\mathring{k}_\pm+\eta)\frac{\partial\vec h}{\partial k}\ ,
\end{equation}
where, by \eqref{eq:k+-eta},
\begin{equation}
\mathring{k}_\pm+\eta=\frac{\displaystyle \Bigl(\frac{\partial h_0}{\partial k}-\phi_k\cdot\frac{\partial h_0}{\partial\phi}\Bigr)\eta}{\displaystyle \frac{\partial h_0}{\partial k}\mp \Bigl|\frac{\partial\vec h}{\partial k}\Bigr|}\ .
\end{equation}

Since
\begin{equation}
\frac{\partial\vec h_\pm}{\partial \phi}=\frac{\partial\vec h}{\partial \phi}+\frac{\partial k}{\partial \phi}\frac{\partial\vec h}{\partial k}\ ,
\end{equation}
\eqref{eq:clocpm}, \eqref{eq:k+-eta}, and \eqref{eq:h+-eta} give
\begin{equation}
C_\text{loc}=\pm\sign\det\Bigl(\frac{\partial\vec h}{\partial{(\phi_1,\phi_2,k)}}\Bigr)\sign\Bigl(\frac{\partial h_0}{\partial k}-\phi_k\cdot\frac{\partial h_0}{\partial\phi}\Bigr)\sign(\eta)\ .
\end{equation}
The result is that, when $\eta$ is swept upwards through the crossing point, the Chern number of the upper (higher $k$) band jumps by
\begin{equation}\label{eq:crossign}
\sign\det\Bigl(\frac{\partial\vec h}{\partial{(\phi_1,\phi_2,k)}}\Bigr)\sign\Bigl(\frac{\partial h_0}{\partial k}-\phi_k\cdot\frac{\partial h_0}{\partial\phi}\Bigr)\ .
\end{equation}

\section{Real scattering matrix}\label{sec:realsca}

\begin{figure}[t]
    \centering
     \begin{subfigure}{.45\textwidth}
        \centering
        \includegraphics[width=\textwidth]{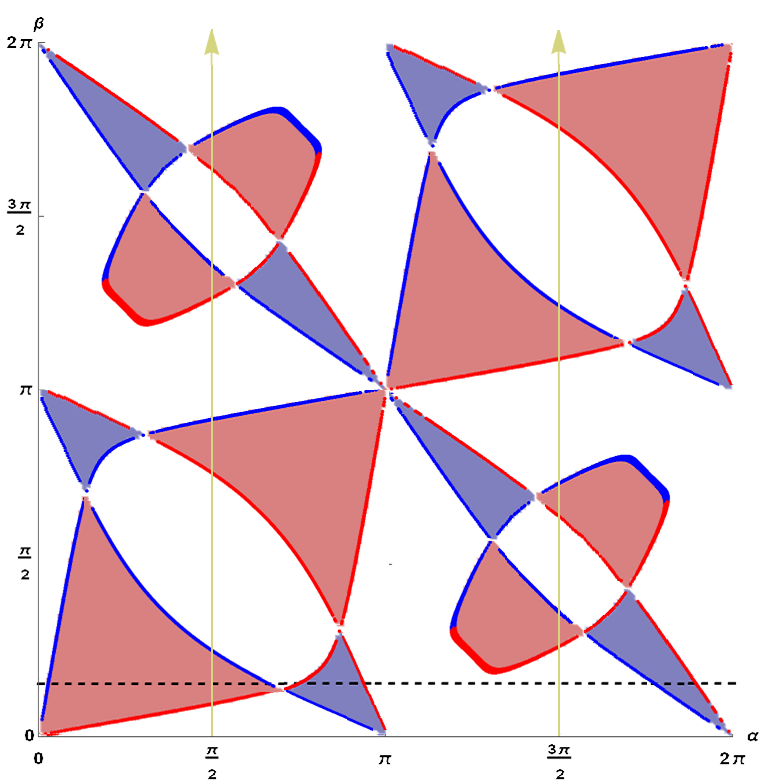}
        \caption{}    
    \end{subfigure}\hspace{1.cm}
\begin{subfigure}{.45\textwidth}
        \centering
        \includegraphics[width=\textwidth]{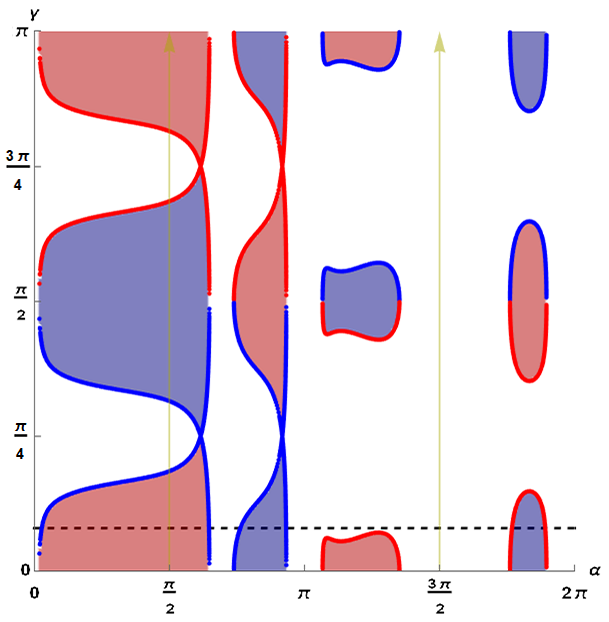}
        \caption{}
        \label{inf_lattice}
    \end{subfigure}% 
    \caption{\label{fig:realS} Topological phase diagram using the same color convention and parametrization scheme as in figure \ref{PD2}.  The scattering matrix  parameters are: $\theta_1 = \pi/4$, $\theta_2 = \pi/3$, $\eta_1 = \pi/2$, $\eta_2 = \pi/6$, and $\nu_1 = \nu_2 =\mu_1=\mu_2= 0$.  In panel (a), we display the Chern number in the  $\alpha\,$--$\,\beta$ plane for  $\gamma=0.25$.  In panel (b), we display the Chern number in the $\alpha\,$--$\,\gamma$ plane for  $\beta=0.5$. The dashed black line corresponds to the same set of parameters $\beta=0.5, \gamma=0.25$ in both panels.}
\end{figure}

\begin{figure}[t]
    \centering
   
    \begin{subfigure}{0.45\textwidth}
        \centering
        \includegraphics[width=\textwidth]{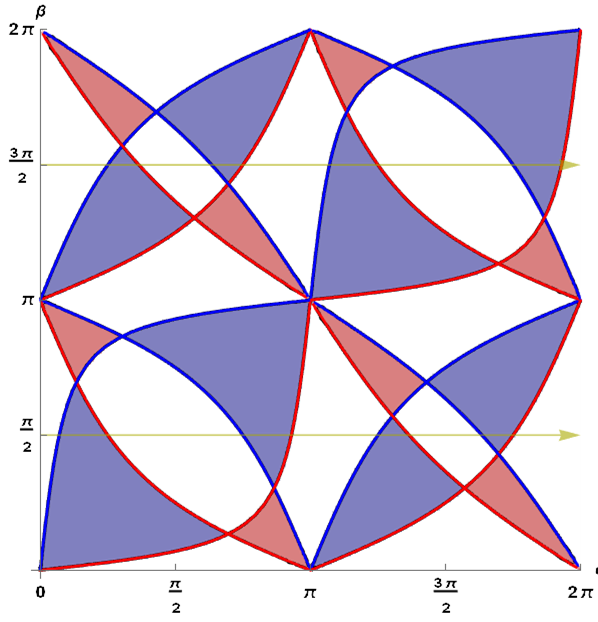}
        \caption{}
        \label{PD1}
    \end{subfigure}
\hspace{1cm}
    \begin{subfigure}{0.45\textwidth}
        \centering
        \includegraphics[width=\textwidth]{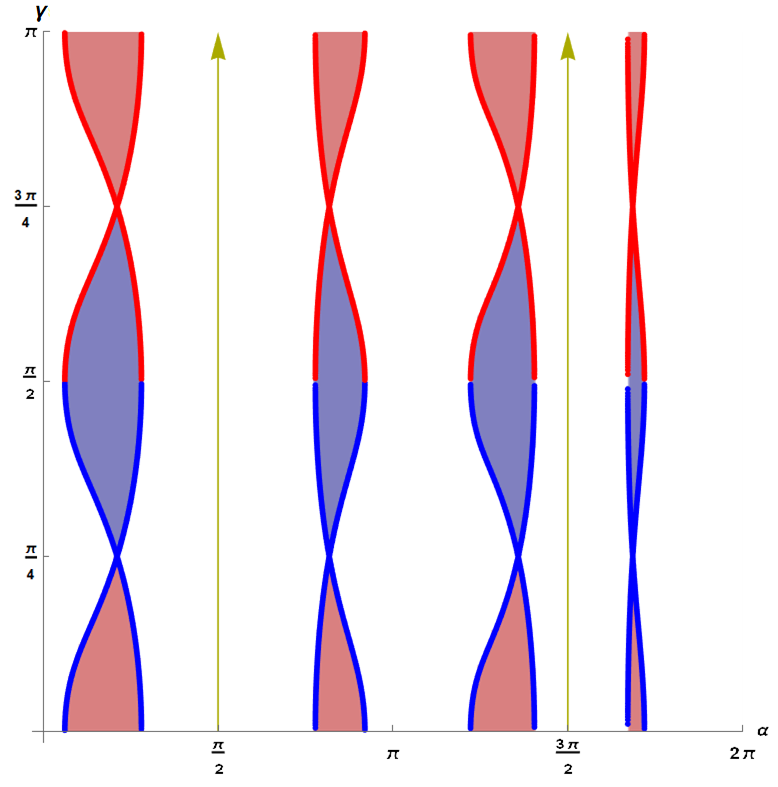}
        \caption{}
        \label{PD3}
    \end{subfigure}
    \caption{ \label{fig:realS2} Topological phase diagram with the same color and parametrization scheme as in figure \ref{PD2}. The parameters are fixed at $\theta_1 = \pi/4$, $\theta_2 = \pi/3$, $\eta_1 = \pi/2$, $\eta_2 = \pi/6$, and $\nu_1 = \nu_2 =\mu_1=\mu_2= 0$. Panel  (a)  shows the Chern number in the $\alpha$--$\beta$ plane for real scattering, $\gamma=0$. Panel (b) shows the Chern number  in the $\alpha\,$--$\,\gamma$ plane for  $\beta=1.44$.
}
\end{figure}

So far we have studied the topological phases of a quantum graph with a general scattering matrix. The results above can be simplified if the matrix $S$ in \eqref{eq:sgamma} is real. In this case, the phases $\delta=\bar\delta=0$ and equation (\ref{gammaSol}) reduces to 
\begin{equation}\label{gammaEQ}
    \gamma=\pm\frac{1}{2}(\psi+\Bar{\psi}).
\end{equation}
The resulting phase diagram is
symmetric under the reflection $\gamma \longleftrightarrow -\gamma$, see figures \ref{fig:realS} and \ref{fig:realS2}.
The intersection between branches occurs when $\psi=-\bar{\psi}\pm\pi/2$ which gives $\gamma=\pm{\pi}/{4}$. For these values of $\gamma$  the Chern number is zero  independently of the choice
of other parameters.

Furthermore, by taking the real and imaginary parts of  (\ref{abcd}) we get 
\begin{equation}
    (a+b)\cos(k+\gamma)=(c+d)\cos\phi_1 \, ,\, (a-b)\sin(k+\gamma)=(c-d)\sin\phi_1.
\end{equation}
From these equations, we have
\begin{equation}
    \bigg(\frac{a+b}{c+d}\bigg)^2\cos^{2}(k+\gamma)+\bigg(\frac{a-b}{c-d}\bigg)^2\sin^{2}(k+\gamma)=1
\end{equation}
and
\begin{equation}\label{EqEPSI}
 \tan(k+\gamma)=\mathcal{E}, \qquad   \mathcal{E}=\pm\left( { -}\,\frac{ \left(\frac{a+b}{c+d}\right)^2-1}{ \left( \frac{a-b}{c-d}\right)^2-1  }\right)^{1/2}\ ;
\end{equation}
a similar set of equations is obtained from (\ref{eq:abcd2}),
\begin{equation}\label{EqEPSI2}
 \tan  (k-\gamma)=\mathcal{\bar E}, \qquad   \mathcal{\bar E}=\pm\left( {-}\, \frac{ \left(\frac{\bar a+\bar b}{\bar c+\bar d}\right)^2-1}{ \left( \frac{\bar a-\bar b}{\bar c-\bar d}\right)^2-1  }\right)^{1/2}.
\end{equation}
Eliminating $k$ between (\ref{EqEPSI}, \ref{EqEPSI2}), the condition for spectral degeneracies can be cast in the form
\begin{align}\label{EqGamma}
    \tan(2\gamma)=\frac{\bar{\mathcal{E}}-\mathcal{E}}{1-\bar{\mathcal{E}}\mathcal{E}}.
\end{align}

The above result can be simplified further in the cases where either  $ \bar{\mathcal{E}}=\mathcal{E}=0$, or $ \bar{\mathcal{E}}=\mathcal{E}=\infty$, so that   $k=0,\pi$ and $k=\pi/2,3\pi/2$, respectively, and $\gamma=0$, so that $S$ is entirely real. In these cases  the band degeneracy conditions reduce to
\begin{equation}
    a+b=\pm(c+d)\,\, ,\,\, a-b=\pm(c-d)\ ,
\end{equation}
which leads to 
\begin{align}\label{pd eq}
    \frac{\cos(\frac{\alpha-\beta}{2})}{\cos(\frac{\alpha+\beta}{2})}=\pm\frac{\cos{\frac{\epsilon_+}{2}}}{\cos{\frac{\theta_+}{2}}} &,   \qquad \frac{\cos(\frac{\alpha-\beta}{2})}{\cos(\frac{\alpha+\beta}{2})}=\pm\frac{\sin{\frac{\epsilon_+}{2}}}{\sin{\frac{\theta_+}{2}}}\\
    \frac{\sin(\frac{\alpha-\beta}{2})}{\sin(\frac{\alpha+\beta}{2})}=\pm\frac{\cos{\frac{\epsilon_-}{2}}}{\cos{\frac{\theta_-}{2}}} &,  \qquad   \frac{\sin(\frac{\alpha-\beta}{2})}{\sin(\frac{\alpha+\beta}{2})}=\pm\frac{\sin{\frac{\epsilon_-}{2}}}{\sin{\frac{\theta_-}{2}}},
\end{align}
where $\theta_+=\theta_1+\eta_1+\theta_2+\eta_2$, $\epsilon_+=-\theta_1-\eta_1+\theta_2+\eta_2$, $\theta_-=\theta_1-\eta_1+\theta_2-\eta_2$ and $\epsilon_-=-\theta_1+\eta_1+\theta_2-\eta_2$, and where $\theta_{1,2},\eta_{1,2}, \alpha,\beta$ are scattering matrix parameters defined in appendix \ref{app:para}.  The  set of equations above explicitly defines    the  boudnary lines   between different  topological phases for a real $S$. 
Equations~(\ref{pd eq})  and (\ref{EqGamma}, \ref{gammaEQ}) were  verified by numerical Chern-number calculations shown in figures \ref{PD1} and \ref{PD3}.

\section{Conclusions}

In this study, we investigated the Chern number  of the  energy bands of a two-dimensional periodic square lattice  quantum graph with a general four-port scatterer at each nodes. This graph is equivalent by a Bloch transformation to a figure-eight graph with two loops pierced by Aharonov-Bohm magnetic fluxes. A direct calculation of Chern numbers is challenging due to the need to find roots of quartic polynomials.
To circumvent this  problem,  we introduced two  open quantum graphs  with reduced  2-by-2 scattering matrices  $\hat S_1$, $\hat S_2$ depending on the magnetic fluxes $\phi_1$ and $\phi_2$, respectively.  

We  showed that there are two possible topological phases: a trivial one with vanishing  Chern numbers for all bands, and  a nontrivial phase, in which the Chern numbers of the bands alternate between $\pm1$. Transitions between the two  topological phases occur precisely when   either  $\hat S_{1}$ or $\hat S_{2}$ is proportional to the identity matrix. At the transition point, the Chern number of each band changes by $\pm 1$, where the sign can be deduced from the derivatives of the matrices  $\hat S_{1,2}$ with respect to the parameters $\phi_1,\phi_2$. This condition enabled us to delineate the boundary in the topological phase diagram between trivial and nontrivial phases, which is a codimension-one submanifold in the space of unitary scattering matrices.

Our analytical results were compared with  numerical calculations of the Chern numbers of the bands, which, unlike the naturally ordered real-valued bands of Hamiltonians, are valued in the unit circle in the complex plane and lack an inherent order. 
We addressed this issue by developing a method to establish and maintain a consistent ordering of the eigenvalues of the scattering matrix throughout our calculations.

It is worth  emphasizing that in  our model the  Chern numbers are restricted to the three values $\pm 1,0$. This is   
 in sharp contrast with  the  quantum-graph lattices in the presence of a magnetic field investigated in  \cite{Goldman2008}, where the range of possible Chern numbers depends on the applied magnetic field. This example suggests that, in general, increasing the complexity  of the unit cell of the quantum graph lattice expands the range of possible Chern numbers.

Finally, although our methodology provides a robust framework for understanding topological phases in general lattice quantum graphs, the analysis in this paper was restricted to the case of equal edge lengths. This assumption generally implies a periodic Chern number structure in the energy bands. Such periodicity is broken in rectangular quantum graph lattices with incommensurate edge lengths in the two directions. The investigation of the resulting Chern number statistics for an infinite number of bands is left for future research.

\section*{Acknowledgments}

This work was funded by the German–Israeli
Foundation within the project Statistical Topology of Complex Quantum Systems, grant number
GIF I-1499-303.7/2019

\begin{appendices}
\section{Parametrization}\label{app:para}
In this paper we use a parametrization of the 4-by-4 unitary scattering matrix $S$ of the quantum graph based on the decomposition
\begin{equation}\label{Sunitarymatrix}
    S=\begin{pmatrix}
       U_1 & 0
        \\
       0 & U_2
    \end{pmatrix}
    \begin{pmatrix}
        D & \Bar{D}\\
        -\Bar{D}&D 
    \end{pmatrix}\begin{pmatrix}
        V_1 & 0\\
        0 & V_2
    \end{pmatrix}
\end{equation}
similar to the one  introduced in \cite{Beenakker1996}.
Here
\begin{equation}
    D=\begin{pmatrix}
        \cos{\alpha}&0\\
        0&\cos{\beta}
    \end{pmatrix}, \qquad \Bar{D}=\begin{pmatrix}
        \sin{\alpha}&0\\
        0&\sin{\beta}
    \end{pmatrix}\ , \end{equation}
$U_1,U_2$ are general two-by-two unitary matrices,
\begin{equation}
    U_i=e^{i\gamma_i}\begin{pmatrix}
        e^{i\xi_i} & 0 \\
        0 & e^{-i\xi_i}
    \end{pmatrix}\begin{pmatrix}
        \cos{\theta_i} & \sin{\theta_i} \\
        -\sin{\theta_i} & \cos{\theta_i}
    \end{pmatrix}\begin{pmatrix}
        e^{i\nu_i} & 0 \\
        0 & e^{-i\nu_i}
    \end{pmatrix}\ ,  
\end{equation}
 and $V_1,V_2$  are two-by-two unitary matrices
\begin{equation}
    V_i=e^{\pm i\sigma}\begin{pmatrix}
        e^{\pm i\mu} & 0 \\
        0 & e^{\mp i\mu}
    \end{pmatrix}\begin{pmatrix}
        \cos{\eta_i} & \sin{\eta_i} \\
        -\sin{\eta_i} & \cos{\eta_i}
    \end{pmatrix}\begin{pmatrix}
        e^{i\zeta_i} & 0 \\
        0 & e^{-i\zeta_i}
    \end{pmatrix}\ ,
\end{equation}
where the upper sign is used for $i=1$ and the lower sign for $i=2$.
The form of the degeneracy conditions (\ref{eq:2deg1}) implies that the band topology depends on the parameters $\alpha,\beta$ of $D$ and on the combination $\gamma=\frac{1}{2}(\gamma_1- \gamma_2)+\sigma$ and the parameters $ \mu,\,\nu_{1,2}, \,\theta_{1,2},\,\eta_{1,2}$ of the matrices $U_i$, $V_i$, while being independent of $\zeta_i$ and $\xi_i$, which can be absorbed into the flux parameters $\phi_1$, $\phi_2$. Thus, altogether,  the Chern numbers depend on $10$ parameters  of the matrix $S$  of the $16$ parameters that define a general four-by-four unitary matrix. Without loss of generality, we set the redundant parameters as   $\zeta_{1,2}=\xi_{1,2} =\sigma=0$ and $\gamma_1 =\gamma=-\gamma_2$.

\section{A second condition for reflectionless scattering}\label{app:chi}
In section \ref{sec:gammak} we derived the condition \eqref{eq:kpsi} for the wave number $k$ at which the two-by-two scattering matrix $\hat S_2$ is reflectionless by eliminating the flux phase $\phi_1$ from equation \eqref{eq:abcd1}. Since $k$ and $\phi_1$ appear in \eqref{eq:abcd1} symmetrically, the same algebra leads to a condition for reflectionless scattering in terms of $\phi_1$, by exchanging $ae^{i\gamma} \leftrightarrow c$, and $be^{-i\gamma} \leftrightarrow d$. 

The result of this calculation is that the condition for reflectionless scattering is 
\begin{equation}
\phi_1=-\gamma+\frac{\delta\pm\tilde\psi}{2}+n\pi 
\end{equation}
with an integer $n$, where
\begin{equation}
\tilde\psi=\pi-\arccos\Bigl(-\frac{|{D}^2-|A|^2-|B|^2|}{2|A||B|}\Bigr)\ ,\quad D=|a|^2-|b|^2
\end{equation}
provided that $|A|,|B|,|D|$ satisfy triangle inequalities, in which case
$\pi-\tilde\psi$ is the angle opposite the side of length $|D|$ in this triangle if $|A|^2+|B|^2>D^2$.

The triangle inequalities for $|A|,|B|,|D|$ are equivalent to the set \eqref{eq:triabc} of $|A|,|B|,|C|$ triangle inequalities, and the two triangles are related by
\begin{equation}
    |A|^2-|B|^2=|C| |D|\ ,\qquad{|C| \sin \tilde\psi}=|D| \sin \psi 
\end{equation}
see figure~(\ref{figtriangle}) for the geometric significance of these parameters.

\section{Numerical calculation}\label{app:numerics}

The Cherns number were calculated numerically using a discretization of the adiabatic parameter space into a grid. The adiabatic potential was computed along links between neighboring points, while the Berry curvature was calculated over closed loops within the grid. This setup allowed us to sum the discretized curvature across the entire parameter space to obtain the Chern number, refining the grid as necessary to capture the variations accurately. Details of the definitions of the adiabatic potential, curvature, and the computation of the Chern number can be found in \cite{Fuk+05}. 

Here, this method was applied to eigenstates of the unitary matrix $\Lambda S$, rather than to those of a Hermitian matrix. For a Hermitian matrix, all eigenvalues are real, so their order is naturally defined. However, for a unitary matrix, the eigenvalues $e^{i k_j}, j=1,\dots, 4$ lie on the unit circle in the complex plane, making it nontrivial to maintain a fixed order of the eigenvalues. We addressed this issue by noting that since $\det(\Lambda S)=1$, $\sum_{j=1}^{4}k_j\equiv 0\pmod {2\pi}$
for any $\phi_1,\phi_2$, so that we can maintain the ordering of levels by adding the appropriate integer multiples of $2\pi$ to the $k_j$ values needed to keep $\sum_{j=1}^{4}\omega_j=0$.
The ordering of levels is retained also during band-crossing events, allowing us to define upper and lower bands near the intersection point.

\end{appendices}

\bibliographystyle{ieeetr}
\bibliography{references}

@article{Pauling1936,
  author  = {Pauling, L.},
  title   = {The diamagnetic anisotropy of aromatic molecules},
  journal = {J. Chem. Phys.},
  volume  = {4},
  pages   = {673},
  year    = {1936}
}

@article{shpiro1982,
  author  = {Shapiro, B.},
  title   = {Renormalization-Group Transformation for the Anderson Transition},
  journal = {Phys. Rev. Lett.},
  volume  = {48},
  pages   = {823},
  year    = {1982}
}

@article{Texier2004,
  author  = {Texier, Ch. and Montambaux, G.},
  title   = {Weak localization in multiterminal networks of diffusive wires},
  journal = {Phys. Rev. Lett.},
  volume  = {92},
  pages   = {186801},
  year    = {2004}
}

@article{slim1997,
  author  = {Kottos, T. and Smilansky, U.},
  title   = {Quantum chaos on graphs},
  journal = {Phys. Rev. Lett.},
  volume  = {79},
  pages   = {4794},
  year    = {1997}
}

@article{Haldane1988,
  author  = {Haldane, F. D. M.},
  title   = {Model for a Quantum Hall Effect without Landau Levels: Condensed-Matter Realization of the "Parity Anomaly"},
  journal = {Phys. Rev. Lett.},
  volume  = {61},
  pages   = {2015},
  year    = {1988}
}

@article{Goldman2008,
  author  = {Goldman, N. and Gaspard, P.},
  title   = {Quantum graphs and the integer quantum Hall effect},
  journal = {Phys. Rev. B},
  volume  = {77},
  pages   = {024302},
  year    = {2008}
}

@article{Beenakker1996,
  author  = {Beenakker, C. W. J.},
  title   = {Random-Matrix Theory of Quantum Transport},
  journal = {Rev. Mod. Phys.},
  volume  = {69},
  pages   = {731},
  year    = {1997}
}

@article{Fuk+05,
  author  = {Fukui, T. and Hatsugai, Y. and Suzuki, H.},
  title   = {Chern Numbers in Discretized Brillouin Zone: Efficient Method of Computing (Spin) Hall Conductances},
  journal = {J. Phys. Soc. Japan},
  volume  = {74},
  pages   = {1674},
  year    = {2005}
}

@article{Chern1946,
  author  = {Chern, S. S.},
  title   = {Characteristic classes of Hermitian Manifolds},
  journal = {Ann. of Math. (2nd Ser.)},
  volume  = {47},
  number  = {1},
  pages   = {85--121},
  year    = {1946}
}

@article{Berry1984,
  author  = {Berry, M. V.},
  title   = {Quantal Phase Factors Accompanying Adiabatic Changes},
  journal = {Proc. R. Soc. A},
  volume  = {392},
  pages   = {45--57},
  year    = {1984}
}

@book{Nakahara2003,
  author    = {Nakahara, Mikio},
  title     = {Geometry, Topology and Physics},
  edition   = {2nd},
  year      = {2003},
  publisher = {Department of Physics, Kinki University},
  address   = {Osaka, Japan}
}

@book{gregoryberkolaikoElementaryIntroductionQuantum2017,
  title = {An Elementary Introduction to Quantum Graphs},
  author = {{Gregory Berkolaiko}},
  year = {2017},
  month = oct,
  volume = {700},
  eprint = {1603.07356},
  primaryclass = {math-ph},
  doi = {10.1090/conm/700},
  urldate = {2025-01-05},
  archiveprefix = {arXiv},
  langid = {english}
}

@article{Band2013,
  author  = {Band, Ram and Berkolaiko, Gregory},
  title   = {Universality of the Momentum Band Density of Periodic Networks},
  journal = {Phys. Rev. Lett.},
  volume  = {111},
  pages   = {130404},
  year    = {2013}
}

@article{Berkolaiko2018,
  author  = {Berkolaiko, Gregory and Comech, Andrew},
  title   = {Symmetry and Dirac points in graphene spectrum},
  journal = {J. Spectr. Theory},
  volume  = {8},
  year    = {2018}
}

@misc{Gnutzmann2008,
  author = {Gnutzmann, Sven and Smilansky, Uzy},
  title  = {Quantum Graphs: Applications to Quantum Chaos and Universal Spectral Statistics},
  year   = {2008},
  note   = {Unpublished manuscript or lecture notes}
}

@article{Sticlet2012,
  author  = {Sticlet, Doru and Pi{\'e}chon, Fr{\'e}d{\'e}ric and Fuchs, Jean-No{\"e}l and Kalugin, Pavel and Simon, Pascal},
  title   = {Geometrical engineering of a two-band Chern insulator in two dimensions with arbitrary topological index},
  journal = {Phys. Rev. B},
  volume  = {85},
  pages   = {165456},
  year    = {2012}
}

@article{avronAdiabaticQuantumTransport1988,
  title = {Adiabatic Quantum Transport in Multiply Connected Systems},
  author = {Avron, J. E. and Raveh, A. and Zur, B.},
  year = {1988},
  month = oct,
  journal = {Reviews of Modern Physics},
  volume = {60},
  number = {4},
  pages = {873--915},
  issn = {0034-6861},
  doi = {10.1103/RevModPhys.60.873},
  urldate = {2022-08-16},
  langid = {english}
}

@article{avronHomotopyQuantizationCondensed1983,
  title = {Homotopy and {{Quantization}} in {{Condensed Matter Physics}}},
  author = {Avron, J. E. and Seiler, R. and Simon, B.},
  year = {1983},
  month = jul,
  journal = {Physical Review Letters},
  volume = {51},
  number = {1},
  pages = {51--53},
  issn = {0031-9007},
  doi = {10.1103/PhysRevLett.51.51},
  urldate = {2025-01-04},
  copyright = {http://link.aps.org/licenses/aps-default-license},
  langid = {english}
}

@article{bernevigQuantumSpinHall2006,
  title = {Quantum {{Spin Hall Effect}} and {{Topological Phase Transition}} in {{HgTe Quantum Wells}}},
  author = {Bernevig, B. Andrei and Hughes, Taylor L. and Zhang, Shou-Cheng},
  year = {2006},
  month = dec,
  journal = {Science},
  volume = {314},
  number = {5806},
  pages = {1757--1761},
  doi = {10.1126/science.1133734},
  urldate = {2022-03-22}
}

@article{braunWindingNumberStatistics2022,
  title = {Winding Number Statistics of a Parametric Chiral Unitary Random Matrix Ensemble},
  author = {Braun, Petr and Hahn, Nico and Waltner, Daniel and Gat, Omri and Guhr, Thomas},
  year = {2022},
  month = may,
  journal = {Journal of Physics A: Mathematical and Theoretical},
  volume = {55},
  number = {22},
  pages = {224011},
  publisher = {IOP Publishing},
  issn = {1751-8121},
  doi = {10.1088/1751-8121/ac66a9},
  urldate = {2022-11-27},
  copyright = {All rights reserved},
  langid = {english}
}

@article{c.w.j.beenakkerRandommatrixTheoryQuantum1997,
  title = {Random-Matrix Theory of Quantum Transport},
  author = {{C. W. J. Beenakker}},
  year = {1997},
  month = jul,
  journal = {Reviews of Modern Physics},
  volume = {69},
  number = {3},
  pages = {731--808},
  issn = {0034-6861, 1539-0756},
  doi = {10.1103/RevModPhys.69.731},
  urldate = {2022-05-22},
  langid = {english}
}

@article{d.j.thoulessQuantizationParticleTransport1983,
  title = {Quantization of Particle Transport},
  author = {{D. J. Thouless}},
  year = {1983},
  month = may,
  journal = {Physical Review B},
  volume = {27},
  number = {10},
  pages = {6083--6087},
  issn = {0163-1829},
  doi = {10.1103/PhysRevB.27.6083},
  urldate = {2020-11-15},
  langid = {english}
}

@article{faureTopologicalChernIndices2000,
  title = {Topological {{Chern Indices}} in {{Molecular Spectra}}},
  author = {Faure, F. and Zhilinskii, B.},
  year = {2000},
  month = jul,
  journal = {Physical Review Letters},
  volume = {85},
  number = {5},
  pages = {960--963},
  issn = {0031-9007, 1079-7114},
  doi = {10.1103/PhysRevLett.85.960},
  urldate = {2024-05-30},
  copyright = {http://link.aps.org/licenses/aps-default-license},
  langid = {english}
}

@article{gatCorrelationsQuantumCurvature2021,
  title = {Correlations of Quantum Curvature and Variance of {{Chern}} Numbers},
  author = {Gat, Omri and Wilkinson, Michael},
  year = {2021},
  month = jun,
  journal = {SciPost Physics},
  volume = {10},
  number = {6},
  pages = {149},
  issn = {2542-4653},
  doi = {10.21468/SciPostPhys.10.6.149},
  urldate = {2021-08-02},
  copyright = {All rights reserved},
  langid = {english}
}

@article{hahnWindingNumberStatistics2023,
  title = {Winding Number Statistics for Chiral Random Matrices: {{Averaging}} Ratios of Parametric Determinants in the Orthogonal Case},
  shorttitle = {Winding Number Statistics for Chiral Random Matrices},
  author = {Hahn, Nico and Kieburg, Mario and Gat, Omri and Guhr, Thomas},
  year = {2023},
  month = nov,
  journal = {Journal of Mathematical Physics},
  volume = {64},
  number = {11},
  pages = {111902},
  issn = {0022-2488},
  doi = {10.1063/5.0164352},
  urldate = {2023-11-05}
}

@article{hahnWindingNumberStatistics2023b,
  title = {Winding Number Statistics for Chiral Random Matrices: {{Averaging}} Ratios of Determinants with Parametric Dependence},
  shorttitle = {Winding Number Statistics for Chiral Random Matrices},
  author = {Hahn, Nico and Kieburg, Mario and Gat, Omri and Guhr, Thomas},
  year = {2023},
  month = feb,
  journal = {Journal of Mathematical Physics},
  volume = {64},
  number = {2},
  pages = {021901},
  issn = {0022-2488},
  doi = {10.1063/5.0112423},
  urldate = {2023-07-06}
}

@article{haldaneModelQuantumHall1988,
  title = {Model for a {{Quantum Hall Effect}} without {{Landau Levels}}: {{Condensed-Matter Realization}} of the "{{Parity Anomaly}}"},
  shorttitle = {Model for a {{Quantum Hall Effect}} without {{Landau Levels}}},
  author = {Haldane, F. D. M.},
  year = {1988},
  month = oct,
  journal = {Physical Review Letters},
  volume = {61},
  number = {18},
  pages = {2015--2018},
  doi = {10.1103/PhysRevLett.61.2015},
  urldate = {2021-07-27}
}

@article{hasanColloquiumTopologicalInsulators2010,
  title = {{\emph{Colloquium}} : {{Topological}} Insulators},
  shorttitle = {{\emph{Colloquium}}},
  author = {Hasan, M. Z. and Kane, C. L.},
  year = {2010},
  month = nov,
  journal = {Reviews of Modern Physics},
  volume = {82},
  number = {4},
  pages = {3045--3067},
  issn = {0034-6861, 1539-0756},
  doi = {10.1103/RevModPhys.82.3045},
  urldate = {2022-09-13},
  langid = {english}
}

@inbook{j.vonneumannBEHAVIOUREIGENVALUESADIABATIC2000,
  title = {{{ON THE BEHAVIOUR OF EIGENVALUES IN ADIABATIC PROCESSES}}},
  booktitle = {World {{Scientific Series}} in 20th {{Century Chemistry}}},
  author = {{J. Von Neumann} and {E. Wigner}},
  year = {2000},
  month = mar,
  volume = {8},
  pages = {25--31},
  publisher = {WORLD SCIENTIFIC},
  doi = {10.1142/9789812795762_0002},
  urldate = {2021-12-30},
  collaborator = {Hettema, Hinne},
  isbn = {978-981-02-2771-5 978-981-279-576-2},
  langid = {english}
}

@article{kaneZ_2TopologicalOrder2005,
  title = {Z\_2 {{Topological Order}} and the {{Quantum Spin Hall Effect}}},
  author = {Kane, C. L. and Mele, E. J.},
  year = {2005},
  month = sep,
  journal = {Physical Review Letters},
  volume = {95},
  number = {14},
  pages = {146802},
  issn = {0031-9007, 1079-7114},
  doi = {10.1103/PhysRevLett.95.146802},
  urldate = {2023-02-24},
  langid = {english}
}

@article{klitzingNewMethodHighAccuracy1980,
  title = {New {{Method}} for {{High-Accuracy Determination}} of the {{Fine-Structure Constant Based}} on {{Quantized Hall Resistance}}},
  author = {v. Klitzing, K. and Dorda, G. and Pepper, M.},
  year = {1980},
  month = aug,
  journal = {Physical Review Letters},
  volume = {45},
  number = {6},
  pages = {494--497},
  publisher = {American Physical Society},
  doi = {10.1103/PhysRevLett.45.494},
  urldate = {2023-07-05}
}

@article{mahitokohmotoTopologicalInvariantQuantization1985,
  title = {Topological {{Invariant}} and the {{Quantization}} of the {{Hall Conductance}}},
  author = {{Mahito Kohmoto}},
  year = {1985},
  journal = {Annals of Physics},
  volume = {160},
  pages = {343--354},
  langid = {english}
}

@article{n.a.sinitsynSemiclassicalTheoriesAnomalous2008,
  title = {Semiclassical Theories of the Anomalous {{Hall}} Effect},
  author = {{N. A. Sinitsyn}},
  year = {2008},
  month = jan,
  journal = {Journal of Physics: Condensed Matter},
  volume = {20},
  number = {2},
  eprint = {0712.0183},
  primaryclass = {cond-mat},
  pages = {023201},
  issn = {0953-8984, 1361-648X},
  doi = {10.1088/0953-8984/20/02/023201},
  urldate = {2025-01-12},
  archiveprefix = {arXiv},
  langid = {english}
}

@article{qiTopologicalInsulatorsSuperconductors2011,
  title = {Topological Insulators and Superconductors},
  author = {Qi, Xiao-Liang and Zhang, Shou-Cheng},
  year = {2011},
  month = oct,
  journal = {Reviews of Modern Physics},
  volume = {83},
  number = {4},
  pages = {1057--1110},
  doi = {10.1103/RevModPhys.83.1057},
  urldate = {2021-07-28}
}

@article{r.b.laughlinQuantizedHallConductivity1981,
  title = {Quantized {{Hall}} Conductivity in Two Dimensions},
  author = {{R. B. Laughlin}},
  year = {1981},
  month = may,
  journal = {Physical Review B},
  volume = {23},
  number = {10},
  pages = {5632--5633},
  issn = {0163-1829},
  doi = {10.1103/PhysRevB.23.5632},
  urldate = {2025-01-04},
  copyright = {http://link.aps.org/licenses/aps-default-license},
  langid = {english}
}

@article{thoulessQuantizedHallConductance1982,
  ids = {thoulessQuantizedHallConductance1982a},
  title = {Quantized {{Hall Conductance}} in a {{Two-Dimensional Periodic Potential}}},
  author = {Thouless, D. J. and Kohmoto, M. and Nightingale, M. P. and {den Nijs}, M.},
  year = {1982},
  month = aug,
  journal = {Physical Review Letters},
  volume = {49},
  number = {6},
  pages = {405--408},
  issn = {0031-9007},
  doi = {10.1103/PhysRevLett.49.405},
  urldate = {2021-01-05},
  langid = {english}
}

@article{walkerUniversalFluctuationsChern1995,
  title = {Universal {{Fluctuations}} of {{Chern Integers}}},
  author = {Walker, Paul N. and Wilkinson, Michael},
  year = {1995},
  month = may,
  journal = {Physical Review Letters},
  volume = {74},
  number = {20},
  pages = {4055--4058},
  issn = {0031-9007, 1079-7114},
  doi = {10.1103/PhysRevLett.74.4055},
  urldate = {2020-08-11},
  langid = {english}
}

@article{swartzbergUniversalChernNumber2023,
  title = {Universal {{Chern}} Number Statistics in Random Matrix Fields},
  author = {Swartzberg, Or and Wilkinson, Michael and Gat, Omri},
  year = {2023},
  month = jul,
  journal = {SciPost Physics},
  volume = {15},
  number = {1},
  pages = {015},
  issn = {2542-4653},
  doi = {10.21468/SciPostPhys.15.1.015},
  urldate = {2024-05-23},
  copyright = {All rights reserved},
  langid = {english}
}

@article{suSolitonsPolyacetylene1979,
  title = {Solitons in {{Polyacetylene}}},
  author = {Su, W. P. and Schrieffer, J. R. and Heeger, A. J.},
  year = {1979},
  month = jun,
  journal = {Physical Review Letters},
  volume = {42},
  number = {25},
  pages = {1698--1701},
  publisher = {American Physical Society},
  doi = {10.1103/PhysRevLett.42.1698},
  urldate = {2025-02-17}
}

@book{leeIntroductionSmoothManifolds2013,
  title = {Introduction to Smooth Manifolds},
  author = {Lee, John M.},
  year = 2013,
  series = {Graduate Texts in Mathematics},
  edition = {2nd ed},
  number = {218},
  publisher = {Springer},
  address = {New York ; London},
  isbn = {978-1-4419-9981-8 978-1-4419-9982-5},
  lccn = {QA613 .L44 2013}
}

@article{von1929no,
  title={No crossing rule},
  author={Von Neumann, John and Wigner, Eugene},
  journal={Z. Phys},
  volume={30},
  pages={467--470},
  year={1929}
}

@article{KottosSmilansky_2003,
doi = {10.1088/0305-4470/36/12/337},
url = {https://doi.org/10.1088/0305-4470/36/12/337},
year = {2003},
month = {mar},
publisher = {},
volume = {36},
number = {12},
pages = {3501},
author = {Tsampikos Kottos and Uzy Smilansky},
title = {Quantum graphs: a simple model for chaotic scattering},
journal = {Journal of Physics A: Mathematical and General}
}

@article{GutkinSmilansky_2001,
doi = {10.1088/0305-4470/34/31/301},
url = {https://doi.org/10.1088/0305-4470/34/31/301},
year = {2001},
month = {jul},
publisher = {},
volume = {34},
number = {31},
pages = {6061},
author = {Boris Gutkin and Uzy Smilansky},
title = {Can one hear the
shape of a  graph?},
journal = {Journal of Physics A: Mathematical and General}
}

@article{GUHR1998,
title = {Random-matrix theories in quantum physics: common concepts},
journal = {Physics Reports},
volume = {299},
number = {4},
pages = {189-425},
year = {1998},
issn = {0370-1573},
doi = {https://doi.org/10.1016/S0370-1573(97)00088-4},
url = {https://www.sciencedirect.com/science/article/pii/S0370157397000884},
author = {Thomas Guhr and Axel Müller–Groeling and Hans A. Weidenmüller}
}

@article{wigner1955,
  title={Characteristic vectors of bordered matrices with infinite dimensions},
  author={Wigner, Eugene P.},
  journal={Annals of Mathematics},
  volume={62},
  number={3},
  pages={548--564},
  year={1955},
  publisher={JSTOR},
  doi={10.2307/1970079}
}

@article{hahnWindingNumberStatistics2025,
  title = {Winding Number Statistics for Chiral Random Matrices: {{Universal}} Correlations and Statistical Moments in the Unitary Case},
  shorttitle = {Winding Number Statistics for Chiral Random Matrices},
  author = {Hahn, Nico and Kieburg, Mario and Gat, Omri and Guhr, Thomas},
  year = 2025,
  month = oct,
  journal = {Journal of Mathematical Physics},
  volume = {66},
  number = {10},
  pages = {101902},
  issn = {0022-2488},
  doi = {10.1063/5.0246969},
  urldate = {2025-10-24}
}

\end{document}